\documentclass[12pt]{article}

\usepackage[text={5.5in,8.5in},centering]{geometry}

\usepackage[title]{appendix}
\usepackage[usenames,dvipsnames,svgnames,table]{xcolor}
\definecolor{citegreen}{rgb}{0.00,0.70,0.30}
\usepackage[colorlinks=true,
            linkcolor=blue,
            urlcolor=blue,
            citecolor=blue,
]{hyperref}
\usepackage{amsrefs}

\usepackage{amssymb}

\usepackage{amsthm}

\usepackage{mathrsfs}
\usepackage{mathtools}

\usepackage{stmaryrd}
\usepackage{gensymb}

\usepackage{xspace}

\DeclareFontFamily{U}{mathx}{\hyphenchar\font45}
\DeclareFontShape{U}{mathx}{m}{n}{
      <5> <6> <7> <8> <9> <10>
      <10.95> <12> <14.4> <17.28> <20.74> <24.88>
      mathx10
      }{}
\DeclareSymbolFont{mathx}{U}{mathx}{m}{n}
\DeclareFontSubstitution{U}{mathx}{m}{n}
\DeclareMathAccent{\widecheck}{0}{mathx}{"71}
\DeclareMathAccent{\wideparen}{0}{mathx}{"75}

\usepackage{pgf}
\usepackage{tikz}

\usetikzlibrary{arrows,automata}

\usetikzlibrary{calc}

\DeclareMathAlphabet{\mathpzc}{OT1}{pzc}{m}{it}

\usepackage{calligra}

\usepackage{epsfig}
\usepackage{enumerate}
\usepackage{graphicx}

\numberwithin{equation}{section}
\newtheorem{theorem}{Theorem}
\newtheorem{prop}{Proposition}[section]

\newtheorem{lemma}{Lemma}[section]

\newtheorem{corollary}{Corollary}[section]

\theoremstyle{remark}
\newtheorem{remark}{Remark}

\theoremstyle{definition}
\newtheorem{definition}{Definition}[section]

\providecommand{\D}{\mathbb}

\providecommand{\R}{\mathrm}

\newcommand{\eu}{\mathrm{e}}

\providecommand{\abs}[1]{\left\lvert#1\right\rvert}

\providecommand{\norm}[1]{\lVert#1\rVert}
\providecommand{\norminf}[1]{\lVert#1\rVert_{\infty}}

\def\ball{\mathrm{B}}

\DeclareMathOperator{\dist}{dist}

\DeclareMathOperator{\Ad}{Ad}

\newcommand{\Spr}[1]{\text{\sc{Spr}}\left[ #1 \right]}
\newcommand{\Stoch}[1]{\csS\left[ #1 \right]}

\DeclareMathOperator{\diag}{diag}
\DeclareMathOperator{\card}{card}
\DeclareMathOperator{\tr}{tr\,}

\DeclareMathOperator*{\supp}{supp}

\DeclareMathOperator{\Unif}{Unif}

\DeclareMathOperator{\one}{\mathbf{1}}

\DeclareMathOperator{\diam}{{\rm diam}}
\DeclareMathOperator{\Mat}{{\text{\rm\sc{Mat}}}}

\def\llb{\llbracket}
\def\rrb{\rrbracket}

\def\mytimesL{\operatornamewithlimits{\hbox{\LARGE$\times$}}}

\def\cdott{\cdot\,}

\newcommand{\mitem}{\par\vskip1mm\noindent$\bullet$\;}

\def\DIV{{\rm\textbf{(DIV)}}\xspace}

\def\UPA{{\rm\textbf{(UPA)}}\xspace}

\def\KKj{{\rm \textbf{K($\mathbf{L_{j}}$)}}\xspace}
\def\KKone{{\rm \textbf{K($\mathbf{L_{j+1}}$)}}\xspace}

\def\lam{{\lambda}}

\def\lea{\lesssim}
\def\gea{\gtrsim}

\def\Balpha{\boldsymbol{\alpha}}
\def\Beps{\boldsymbol{\eps}}

\def\Bfeta{\boldsymbol{\eta}}

\def\Bxi{\boldsymbol{\xi}}

\def\tom{\tilde{\omega}}
\def\th{\vartheta}
\def\Th{{\mathrm{\Theta}}}
\def\tTh{{\mathrm{\widetilde{\Theta}}}}

\def\eps{\epsilon}
\def\ups{{\sigma}}

\def\veps{\varepsilon}

\def\kap{\kappa}

\newcommand{\mchi}{{\chi}}

\def\lamj{\lam^{j}}
\def\lamz{\lam^{0}}

\def\tlamjo{\tilde{\lam}^{j+1}}

\def\lami{\lam^{i}}
\def\lamione{\lam^{i+1}}

\def\Uj{\mathrm{U}^{j}}
\newcommand\Ujinv{(\mathrm{U}^{j})^{-1}}
\newcommand\UjT{(\mathrm{U}^{j})^{\top}}

\def\tUjo{\widetilde{\mathrm{U}}^{j+1}}

\def\Ui{\mathrm{U}^{i}}

\def\Wjo{\mathrm{W}^{j+1}}

\def\Zjo{\mathrm{Z}^{j+1}}

\newcommand{\Sz}{\mathrm{S}^{0}}
\newcommand{\Si}{\mathrm{S}^{i}}

\newcommand{\Sj}{\mathrm{S}^{j}}

\def\Dz{\rD^{0}}
\def\Dj{\rD^{j}}

\newcommand\tDj{\tilde \rD^{j}}

\newcommand\tDi{\tilde \rD^{i}}

\def\Qj{\rQ^{j}}
\def\tQz{\widetilde{\rQ}^{0}}
\def\tQj{\widetilde{\rQ}^{j}}

\def\Cj{\rC^{j}}

\def\Fj{\mathrm{F}^{j}}
\def\Fjo{\mathrm{F}^{j+1}}

\def\Fi{\mathrm{F}^{i}}

\newcommand{\Mi}{\mathrm{M}^{i}}

\def\Mjo{\mathrm{M}^{j+1}}

\def\psij{\psi^{j}}
\def\psiz{\psi^{0}}

\def\phij{\varphi^{j}}
\def\phiz{\varphi^{0}}

\def\tffijo{\widetilde{\varphi}^{j+1}}

\def\psii{\psi^{i}}

\def\phii{\varphi^{i}}
\def\phiione{\varphi^{i+1}}
\def\ffii{\varphi^{i}}

\def\Psii{\Psi^{i}}
\def\Psij{\Psi^{j}}
\def\Psijo{\Psi^{j+1}}

\def\Lamj{\Lambda^{j}}

\def\Lamjo{\Lambda^{j+1}}

\def\Lami{\Lambda^{i}}

\def\epsj{\eps_{j}}

\def\lamjo{\lambda^{j+1}}

\def\tlamjo{\tilde{\lambda}^{j+1}}

\def\psijo{\psi^{j+1}}
\def\phijo{\varphi^{j+1}}

\def\ffijo{\varphi^{j+1}}

\def\ord#1{o\left[ #1 \right]}

\def\Ord#1{O\left[ #1 \right]}

\def\bigOrd#1{O\big[ #1 \big]}

\def\tV{{ \widetilde{V} }}

\def\hk{{\hat{k}}}

\def\hth{{{\widehat{\th}}}}

\def\BM{\mathbf{M}}

\newcommand{\Bc}{\mathbf{c}}

\def\Bt{\mathbf{t}}

\def\BLam{\mathbf{\Lambda}}

\def\Const{{\rm{Const\,}}}

\def\tn{{\tilde{n}}}

\def\DC{\D{C}}

\def\DP{\D{P}}

\def\DR{\D{R}}
\def\DT{\D{T}}
\def\DZ{\D{Z}}
\def\DN{\D{N}}

\def\cA{\mathcal{A}}
\def\cB{\mathcal{B}}

\def\csL{\mathscr{L}}

\def\csS{\mathscr{S}}

\def\cC{\mathcal{C}}

\def\cE{\mathcal{E}}

\def\cH{\mathcal{H}}

\def\cL{\mathcal{L}}

\def\cP{\mathcal{P}}

\def\cQ{\mathcal{Q}}

\def\cT{\mathcal{T}}

\def\cX{\mathcal{X}}

\def\cA{{\mathcal{A}}}

\def\rAut{{\R{Aut}}}

\def\rC{{\R{C}}}
\def\rD{{\R{D}}}

\def\rd{{\R{d}}}

\def\rx{{\R{x}}}
\def\ry{{\R{y}}}

\def\rP{{\R{P}}}
\def\rQ{{\R{Q}}}

\def\rS{{\R{S}}}

\newcommand{\hj}{\hat{\mathrm{j}}}

\newcommand{\hn}{{\hat{\mathrm{n}}}}

\newcommand{\lcite}[2]{{\cite[#2]{#1}}}

\def\be{\begin{equation}}
\def\ee{\end{equation}}
\def\ba{\begin{array}{l}}
\def\ea{\end{array}}
\def\bal{\begin{aligned}}
\def\eal{\end{aligned}}

\newcommand{\beal}{\begin{equation}\begin{aligned}}
\newcommand{\eeal}{\end{aligned}\end{equation}}

\def\bea#1\eea{\begin{align}#1\end{align}}
\def\bea#1\eea{\begin{align}#1\end{align}}

\def\ble{\begin{lemma}}
\def\ele{\end{lemma}}

\def\bco{\begin{cor}}
\def\eco{\end{cor}}

\def\bpr{\begin{prop}}
\def\epr{\end{prop}}

\def\bre{\begin{remark}}
\def\ere{\end{remark}}

\def\btm{\begin{theorem}}
\def\etm{\end{theorem}}

\def\fB{\mathfrak{B}}
\def\fD{\mathfrak{D}}
\def\fF{\mathfrak{F}}

\def\fa{\mathfrak{a}}

\def\ff{\mathfrak{f}}
\def\fg{\mathfrak{g}}

\def\fs{\mathfrak{s}}

\def\om{{\omega}}
\def\Om{{\Omega}}

\def\eps{\epsilon}

\def\Lam{{\Lambda}}
\def\lam{{\lambda}}

\def\thnk{{\th_{n,k}}}

\def\mchink{{\mchi_{n,k}}}

\def\Cnk{{C_{n,k}}}

\def\Thj{{\Th^{j}}}

\def\Thi{{\Th^{i}}}
\def\tThi{{\tTh^{i}}}

\def\Thinf{{\Th^{\infty}}}

\def\DPTh{\DP^{\Theta}}

\def\cond{\,\big|\,}

\def\prth#1{{\DP^{\Th}\left\{\,#1\,\right\}}}

\def\prthp{{\DP^{\Th}}}

\def\esm#1{\D{E}\left[\, #1\, \right]}
\def\pt{\partial}
\def\half{\frac{1}{2}}
\def\shalf{{\textstyle{\frac{1}{2}}}}

\def\quart{\frac{1}{4}}
\def\squart{{\textstyle{\frac{1}{4}}}}

\def\tto#1{\smash{\mathop{\,\,\,\, \longrightarrow \,\,\,\, }\limits_{#1}}}

\newcommand{\all}{\forall\,}

\newcommand{\monb}{}

\newcommand{\redtwo}{{\rrr{\mathbf{2}}}}

\def\be{\begin{equation}}
\def\ee{\end{equation}}
\def\ba{\begin{array}{l}}
\def\ea{\end{array}}
\def\bal{\begin{aligned}}
\def\eal{\end{aligned}}

\def\ble{\begin{lemma}}
\def\ele{\end{lemma}}

\def\bre{\begin{remark}}
\def\ere{\end{remark}}

\def\btm{\begin{theorem}}
\def\etm{\end{theorem}}

\def\bde{\begin{definition}}
\def\ede{\end{definition}}

\def\bpr{\begin{prop}}
\def\epr{\end{prop}}

\def\bco{\begin{corollary}}
\def\eco{\end{corollary}}

\definecolor{redd}{rgb}{0.95,0.2,0.2}
\definecolor{gris}{rgb}{0.9,0.9,0.9}

\definecolor{cmgray}{rgb}{0.7,0.7,0.7}

\definecolor{greenn}{rgb}{0.1,0.6,0.2}

\def\tmgreen#1{\textcolor{greenn}{#1}}

\definecolor{cmblue}{rgb}{0.2,0.5,0.8}

\DeclareMathAlphabet{\mathcalligra}{T1}{calligra}{m}{n}

\definecolor{whitte}{rgb}{1.0,1.0,1.0}
\definecolor{redd}{rgb}{0.95,0.2,0.2}
\definecolor{gris}{rgb}{0.9,0.9,0.9}
\definecolor{grisfive}{rgb}{0.5,0.5,0.5}
\definecolor{greenn}{rgb}{0.1,0.6,0.2}

\definecolor{citegreen}{rgb}{0.00,0.70,0.30}

\def\rrr#1{\textcolor{redd}{#1}}

\def\ggggg#1{\textcolor{gris}{ {\tmgreen{[[}} HIDDEN {\tmgreen{]]}}}}

\definecolor{redd}{rgb}{0.95,0.2,0.2}
\definecolor{gris}{rgb}{0.9,0.9,0.9}

\definecolor{cmgray}{rgb}{0.7,0.7,0.7}

\definecolor{greenn}{rgb}{0.1,0.6,0.2}

\def\tmgreen#1{\textcolor{greenn}{#1}}

\definecolor{cmblue}{rgb}{0.2,0.5,0.8}

\def\bs{\begin{equation*}}
\def\es{\end{equation*}}

\def\bl{\begin{equation}\begin{aligned}}
\def\el{\end{aligned}\end{equation}}

\def\bls{\begin{equation*}\begin{aligned}}
\def\els{\end{aligned}\end{equation*}}

\def\Uz{\mathrm{U}^{0}}

\def\Lamz{\Lambda^{0}}
\def\Lamo{\Lambda^{1}}

\newcommand{\vertiii}[1]{{\vert\kern-0.30ex\vert\kern-0.30ex\vert #1 \vert\kern-0.30ex\vert\kern-0.30ex\vert}}
\newcommand{\bvertiii}[1]{{\big\vert\kern-0.50ex\big\vert\kern-0.50ex\big\vert #1 \big\vert\kern-0.50ex\big\vert\kern-0.50ex\big\vert}}
\newcommand{\Bvertiii}[1]{{\Big\vert\kern-0.40ex\Big\vert\kern-0.40ex\Big\vert #1
    \Big\vert\kern-0.40ex\Big\vert\kern-0.40ex\Big\vert}}

\providecommand{\mxnorm}[1]{\norm{#1}_{x}}

\providecommand{\vmnorm}[1]{\vertiii{#1}}

\newcommand{\ptt}{\partial_{t}}

\newcommand{\pttz}{\partial_{t_z}}

\def\Fz{\mathrm{F}^{0}}
\def\Fo{\mathrm{F}^{1}}

\def\psiz{\psi^{0}}
\def\psio{\psi^{1}}

\def\lamz{\lam^{0}}
\def\lamo{\lam^{1}}

\def\Psiz{\Psi^{0}}

\def\Qz{\mathrm{Q}^{0}}

\def\Mo{\mathrm{M}^{1}}

\def\Wo{\mathrm{W}^{1}}

\newcommand{\para}[1]{\left(#1\right)}
\newcommand{\brak}[1]{\left[#1\right]}

\newcommand{\bpara}[1]{\big(#1\big)}

\newcommand{\Bpara}[1]{\Big(#1\Big)}
\newcommand{\Bbrak}[1]{\Big[#1\Big]}

\newcommand{\set}[1]{\left\{#1\right\}}
\newcommand{\mins}{\setminus}

\newcommand{\zplus}{{0^{\scalebox{0.5}{$+$}}}}

\newcommand{\onemi}{{1^{\scalebox{0.5}{$-$}}}}
\newcommand{\twomi}{{2^{\scalebox{0.5}{$-$}}}}
\newcommand{\mmi}[1]{{#1^{\scalebox{0.5}{$-$}}}}
\newcommand{\mpi}[1]{{#1^{\scalebox{0.5}{$+$}}}}
\newcommand{\mpmi}[1]{{#1^{\scalebox{0.5}{$\pm$}}}}

\def\tffi{\widetilde{\varphi}}

\def\tffio{\widetilde{\varphi}^{1}}

\def\llb{\llbracket}
\def\rrb{\rrbracket}
\newcommand{\lrb}[1]{\llbracket #1 \rrbracket}

\newcommand{\nblz}{\vskip1mm\noindent$\blacklozenge\,$}

\providecommand{\bigabs}[1]{\big\lvert#1\big\rvert}
\providecommand{\Bigabs}[1]{\Big\lvert#1\Big\rvert}

\newcommand{\vide}{\varnothing}

\newcommand{\msperp}{{}}

\def\DX{\mathbb{X}}
\def\DY{\mathbb{Y}}
\newcommand{\normX}[1]{\norm{#1}_{\DX}}
\newcommand{\normY}[1]{\norm{#1}_{\DY}}
\def\xball{\cB}
\newcommand{\mkap}{\kappa}
\def\ballcA{\cB^{\cA}}

\begin{document}

\title{Non-local Minami-type estimates\\for a class of quasi-periodic  media}

\author{Victor Chulaevsky}



\maketitle

\centerline{\small Universit\'{e} de Reims, D\'{e}partement de Math\'{e}matiques}
\centerline{\small  Moulin de la Housse, B.P. 1039}
\centerline{\small 51687 Reims Cedex 2, France}
\centerline{\small E-mail: victor.tchoulaevski@univ-reims.fr}
\vspace{1mm}

\begin{abstract}
This paper is a follow-up of our earlier work \cite{C18b} where a uniform exponential Anderson localization was proved for a class of deterministic (including quasi-periodic)
Hamiltonians with the help of a variant of the KAM (Kolmogorov--Arnold--Moser) approach. Building on \cite{C18b},
we prove for the same class of operators a non-local variant of the Minami eigenvalue concentration estimate.
\end{abstract}

\maketitle

\section{Introduction}
\label{intro}

We study spectral properties of finite-difference operators, usually called
lattice Schr\"{o}dinger operators (LSO), acting in the Hilbert space
$\cH = \ell^2(\DZ^d)$,
\bl
\label{eq:def.H}
\bpara{ H_\veps(\om,\th) f }(x) &= \veps \bpara{ \Delta f }(x)  + v(T^x\om,\th) f(x),\; x,y\in \DZ^d, \, g\in\DR,
\\
\bpara{ \Delta f }(x) &= \sum_{y:\, |y-x|=1} f(y)\,,
\el
where

\mitem $\om\in\Om := \DT^\nu = \DR^\nu/\DZ^\nu \cong [0,1)^\nu$, $\nu\ge 1$;

\mitem $T:\DZ^\nu\times\Om\to\Om$ is a conservative dynamical system;

\mitem $\th\in\Th$, where $(\Th,\fB,\prthp)$ is an auxiliary probability space.

The amplitude $\veps>0$ of the kinetic energy operator is assumed to be small.
The dynamical system $T$ leaves $\th\in\Th$ invariant, so the latter labels the
operator ensembles $\{H_\veps(\om,\th), \, \om\in\Om\}$. The function
$v:\Om\to\DR$, which we call the hull of the deterministic potential $V$,
has the same general form as in \cites{C14b,C18b}, viz.
\be
\label{eq:def.v}
v(\om, \th) = \sum_{n=0}^\infty a_n \sum_{k=1}^{K_n} \th_{n,k} \mchi_{n,k}(\om),
\ee
but, unlike \cite{C18b} where $\mchi_{n,k}$ were the Haar wavelets on the torus, we work now with non-orthogonal functions $\mchink$.
This provides some notational simplifications, which seem to be welcome as the presentation here is quite technical, but our methods
and results extend easily to the model with Haar wavelets. For clarity, we assume that $\Om = \DT^\nu = (\DR/\DZ)^\nu$, and set $\mchink = \one_{\Cnk}$, where
\bl
\label{eq:def.C.n.k}
\Cnk = \mytimesL_{j=1}^\nu \big[\, l_j(k) 2^{-n}, (l_j(k)+1)2^{-n}  \,\big) \,, \quad n\ge 1,
\el
are cubes, generating a partition of $\DT^\nu$ which we denote by $\cC_n$. Within $\cC_n$, the cubes $C_{n,k}$ are numbered in some way,
and the ordinal number $k \in \lrb{1, K_n}$, with $K_n := 2^{\nu n}$, of such a cube determines a unique multi-index
$\bpara{ l_1(k), \ldots \,, l_\nu(k) } \in \lrb{0, 2^n - 1}^\nu$. Here and below, $\lrb{a,b} := [a,b] \cap \DZ$.

The factors $\{\thnk, n\ge 1, \; 1 \le k \le K_n \}$  form a family of IID (independent and identically distributed) random variables on $\Th$
uniformly distributed in $[0,1]$.
The motivation for such a model can be found in \cites{C12,C14b}.

The two most frequently used approaches to the Anderson localization phenomena (except for one-dimensional models; cf. \cites{GMP77,KS80,Carm82})
are the multi-scale analysis (MSA, cf. \cites{FS83,FMSS85,Dr87,Sp88,DK89})
and the fractional moments method (FMM, cf. \cites{AM93,ASFH01}). The latter, when applicable, usually provides a simple, quite transparent proof and strong (exponential)
bounds on the decay of the eigenfunction correlators, but, unfortunately, does not apply to the deterministic operators, for it requires fully developed disorder;
ideally, one needs an IID random potential. The MSA, on the contrary, is much more flexible (cf., e.g., \cite{Sp88}), but the localization estimates it can provide are slightly
weaker.

The KAM (Kolmogorov--Arnold--Moser) type technique constitutes an alternative to MSA and FMM. Historically, this approach to the Anderson localization
pre-dates both MSA and FMM; see  \cite{BLS83} where the eigenbasis of a quasi-periodic Hamiltonian was constructed recursively, by \textbf{unitary} transformations.
In the present paper, we use a different variant of the KAM technique proposed in \cite{Sinai87}, and work with a sequence of approximate eigenbases which are only
\textbf{approximately} orthogonal, with precision improving in the course of an inductive procedure.
The class of deterministic (including quasi-periodic) operators we address here is the same as in \cite{C18b} where a very strong form
of localization, usually called ULE  (Uniform Localization of Eigenfunctions) was proved, along with the uniform unimodality of the eigenfunctions.

A more detailed discussion of prior results and alternative techniques can be found in \cite{C14b};
we refer the reader to the works \cites{Jit99,BG00,BGS01,Chan07,DamG11,DamG13}.

$\blacklozenge$ In this paper, we address the problem of regularity of correlation measures of a finite family of eigenvalues $\lam_1(\om)$, \ldots \, $\lam_K(\om)$, $K\ge 1$,
numerated in some measurable way. The pioneering work by Wegner \cite{W81} established Lipschitz-continuity of the correlation measure for $K=1$, usually called
integrated density of states, in the lattice Anderson model with IID potentials $V(x,\om)$ under the assumption of Lipschitz continuity of the common marginal measure.
Surprisingly, it took fifteen more years to treat a particular case for $K=2$.
{The original Minami estimate \cite{Min96} for $K=2$ eigenvalues was extended to any $K \ge 2$ in \cites{BHS07,GV07} and refined in several subsequent works.
A particularity of the aforementioned works consists in the fact that the regularity of the the two-point correlation measure, depending of course upon
the regularity of the IID random potential, was proved for the squares $I\times I\subset \DR$ and not for arbitrary rectangles $I_1\times I_2$.
It is to be noted that a weaker result was proved in \cite{CGK09} for rectangles $I_1\times I_2$, assuming the random potential is IID with continuity modulus
$\fg(\cdot)$ of the common marginal measure:
\bl
\notag
&
{\esm{ \para{ \tr P^\Lam_{I_1} \para {H(\om) } } \, \tr P^\Lam_{I_2}\para{ H(\om) }
    - \min \brak{ \tr P^\Lam_{I_1}\para{ H(\om) } \,, \tr P^\Lam_{I_2}\para{ H(\om) } } }  }
\\
& \qquad
\le 2 \, |\Lam|^2 \fg_\Lam( |I_1|) \, \fg_\Lam( |I_2|) \,,
\el
while a more efficient estimate
\bl
\label{eq:CGK.subset}
\esm{ \para{ \tr P^\Lam_{I_1}(H(\om)) } \, \tr P^\Lam_{I_2}(H(\om))  } \le  |\Lam|^2 \fg_\Lam( |I_1|) \, \fg_\Lam( |I_2|)
\el
was proved only in the case where $I_1 \subset I_2$.
See further references  in \cite{CGK09}.

The general case where the positions of the intervals $I_1, I_2$ can be arbitrary is known to be more difficult.
Yet, there are situations where one needs to allow for the intervals $I_1, I_2$ not necessarily close to each other.
The main motivation for the present work is an application to the N-body localization problems, so the results obtained here
will be used in a forthcoming work \cite{CS22atmp} on N-body localization in a quasi-periodic potential of the form \eqref{eq:def.v}.

In our model, one can only make use of a parametric "randomness", but once a hull function $v:\, \Om = \DT^\nu\to \DR$ is fixed,
the joint probability measure of the eigenvalues, say, in a finite cube, is quite singular; this is why we do not make any use of it.
Instead, we establish the key properties of the Hamiltonian $H(\om,\th)$ for all (and not just almost all) $\om\in\Om$.

The aforementioned parametric randomness or, better to say, freedom is insufficient to prove even a Wegner-type estimate, let alone Minami estimates, with an
optimal volume dependence.
However, in applications to the Anderson localization problems, one rarely, if ever, needs optimal Wegner/Minami bounds (cf., e.g., \cite{KleMol06}).
Thus we settle for a weaker estimate but for arbitrary pairs of intervals $I_1, I_2$, with
a large volume-dependent factor in the RHS of \eqref{eq:CGK.subset}: $|\Lam|^C$ with some $C>2$. Optimality is achieved, however,
in dependence upon $|I_{1,2}|$: our bound is bilinear in $(|I_1|, |I_2|)$.

\subsection{Requirements for the dynamical system}\label{ssec:UPA}

We consider only the case where the phase space of the underlying dynamical system is $\Om = \DT^\nu$, $\nu\ge 1$,
and endow $\Om$ with the distance $\dist_\Om(\cdot\,, \cdot)$
inherited from the max-distance in $\DR^\nu$:
$$
\dist_\Om(\om', \om'')
:= \max_{1 \le i \le \nu} \dist_{\DT^1}(\om'_i, \om''_i) .
$$
We assume that the dynamical system $T:\, \DZ^d\times\Om \to \Om$ fulfills the condition of uniform power-law aperiodicity
\par\vskip1mm\noindent
\UPA \quad $\exists\, A, C_A\in\DN^* \; \;\forall\, \om\in\Om \;
  \forall\, x, y\in\DZ^\nu \text{ with } x\ne y $
\be
\label{eq:cond.UPA}
\notag
\begin{array}{lc}
\quad \dist_{\Om}(T^x \om, T^y \om) \ge C^{-1}_A |x - y |^{-A},
\end{array}
\ee
and of tempered divergence of trajectories:
\par\vskip1mm\noindent
\DIV \quad $\exists\, A', C_{A'}\in\DN^* \; \;\forall\, \om, \om'\in\Om \; \forall\, x\in\DZ^\nu\mins\set{0}$
\be
\notag
\begin{array}{lc}
\quad \dist_{\Om}(T^x \om, T^x \om') \le C_{A'} |x|^{A'} \dist_{\Om}(\om, \om').
\end{array}
\ee
For the toral shifts $T^x\om = \om + x_1 \Balpha_1 + \cdots + x_d \Balpha_d$, $\Balpha_i\in\DT^\nu$, \UPA
is a Diophantine condition on the frequency vectors $\Balpha_i$, while \DIV
is trivially fulfilled, since $\set{T^x\,, x\in\DZ^d}$ are isometries of the torus.

\subsection{Main results}
\label{ssec:Main.res}

The results on localization and generalized Minami-type estimates presented below apply to lattice Schr\"{o}dinger operators with
deterministic potentials not only in the entire lattice $\DZ^d$, but also to their restrictions to some lattice subsets
with a sufficiently simple boundary, first of all to finite cubes and, more generally, to finite intersections of
half-lattices of the form $\set{x\in\DZ^d:\, s_i x_i \ge a_i}$, $s_i \in\set{-1,\, 1}$, $a_i\in\DZ$. For brevity, we call such subsets
$\cQ \subset \DZ^d$ \emph{simple}. In fact, our methods apply to a larger class of subsets; we comment on it later.

Except for the extension to proper subsets of $\DZ^d$ (cf. Appendix \ref{app:loc.simple.sets}) and some minor modifications of the scaling parameters,
Theorem \ref{thm:main.loc} is proved as in \cite{C18b}, but the proof of the new result, Theorem \ref{thm:Main},
heavily relies on the inductive construction of the eigenvalues and eigenfunctions, so the essential steps of the proof from \cite{C18b} are reproduced
in Section \ref{sec:KAM}.

\btm\label{thm:main.loc}
Consider the family of lattice Schr\"{o}dinger operators \eqref{eq:def.H}.
Under the hypotheses \UPA and \DIV, there exists $\veps_* \in(0,+\infty)$ such that, for any $\veps\in (0, \veps_*)$,
there exists a subset $\Thinf(\veps) \subset \Th$ with $\prth{\Thinf(\veps)} \uparrow 1$ as $\veps\downarrow 0$
and with the following property: if $\th\in\Thinf(\veps)$, then for \textbf{any} $\om\in\Om$:
\begin{enumerate}[\rm(A)]
  \item $H_\veps(\om,\th)$ has pure point spectrum;

  \item for any $x\in\DZ^d$, there is exactly one eigenfunction $\varphi_x(\cdott;\om;\th)$ such that
$$
|\varphi_x(x;\om,\th)|^2 > 1/2,
$$
  i.e., $\varphi_x$ has the ``localization center'' $x$, so there is a natural bijection between the elements of the eigenbasis $\{\varphi_x(\cdot;\om,\th)\}$
  and the lattice $\DZ^d$;

  \item
the eigenfunctions $\varphi_x$ decay uniformly away from the localization centers:
\bl
\label{eq:main.loc.unimodal}
\forall\,  y\in\DZ^d\;\; |\varphi_x(y;\om,\th)| \le  \eu^{-m |y-x|}, \;
m = m(\veps) \tto{\veps\to 0} +\infty.
\el
\end{enumerate}
\etm

The simplicity of spectrum in our model was established in \cite{C14b}.

In Appendix \ref{app:loc.simple.sets}, we explain how to adapt the proof of Theorem to the restrictions of $H_\veps$ to certain subsets $\cQ\subsetneq \DZ^d$
(which we call \emph{simple}), including all rectangles $\lrb{a_1, b_1} \times \cdots \times \lrb{a_d, b_d}$;
here and below, $\lrb{a,b} \equiv [a,b]\cap \DZ$. In fact, this remark could have been made already in \cite{C18b}.
The main issue here is the loss of the global covariance property of the spectral data.

\btm
\label{thm:Main}
Let $\cQ \subseteq \DZ^d$ be a simple lattice set, and pick any finite subset
$\cX = \set{ x_1\,, \ldots,  x_N }\subset \cQ^N$ with $\card \cX = N$.
Under the assumptions and with notations of Theorem \ref{thm:main.loc},
for any bounded intervals $I_1, \ldots \,, I_N \subset \DR$, the eigenvalues $\lam_{x_1}\,, \ldots, \lam_{x_N}$
admit the following bound:
\bl
\notag
\prth{ \all k\in\lrb{1,N} \;\; \lam_{x_k} \in I_k }
&
\le  \eu^{ cN\ln^2 (\diam (\cX)) } \prod_{1\le k \le N} \abs{I_k} \,.
\el
\etm

\section{Phase space analysis and spectral spacings}
\label{sec:potential}

Similar to the works \cites{BLS83,C14b}, we establish a complete localization of the eigenfunctions
of $H_\veps(\om,\th)$ for every (and not just almost every) phase point $\om\in\Om \equiv \DT^\nu$.
As stated in the inductive
hypothesis \KKj (cf. Section \ref{sec:KAM}), each induction step can be carried out for \emph{all}
$\om\in\Om$ but only outside a subset of $\Th$ of small $\DPTh$-measure: the smaller $\eps>0$, the
smaller is the measure of the excluded subset. In other words, for $\DPTh$-a.e. $\th\in\Th$
there exists $\eps_\circ(\th)>0$ such that for $\eps\in\big(0,\eps_\circ(\th)\big)$ the operator ensemble
$H_\veps(\cdot,\th)$ on the phase space of the ergodic dynamical system $T$ features a \emph{uniform}
(and not just semi-uniform) complete Anderson localization with unimodal (cf. \eqref{eq:main.loc.unimodal}) eigenfunctions.

In this section, we prepare the ground for the main measure-theoretic estimate (in the parameter
space $\Th$) required for the KAM induction.

\bde
Let be given a dynamical system $T:\DZ^d\times\Om\to\Om$, an arbitrary set $A$,
and an action $\rS$ of the abelian group $\DZ^d$ on $A$, i.e., a homomorphism
$\rS: \, \DZ^d \to \rAut(A)$ of $\DZ^d$ into the group of transformations of $A$.
A mapping $F: \DZ^d \times \Om \to A$ is called $T$-covariant iff
\bl
\notag
\forall\, x\in\DZ^d \qquad F(x,  \om) = \rS^x F(0, T^x\om).
\el
\ede

We shall need three kinds of covariant mappings:

\begin{enumerate}[(i)]
  \item scalar functions $\lam:\, (x,\om) \mapsto \lam_x(\om)\in\DC$, satisfying
$\lam_x(\om) = \lam_0(T^x \om)$;
  \item vector-valued mappings $f:\,(x,\om)\mapsto f_x(\cdot, \om)$ with values in $\ell^2(\DZ^d)$,
  so that a function $y \mapsto f_x(y,\om)$ with fixed $x$ and $\om$ is square-summable and satisfies
$f_x(y, \om) = f_0(y-x, T^x \om)$;

  \item matrix-valued mappings $F:\DZ^d\times\Om\to\Mat(\DZ^d)$   constant in the first argument, so that its matrix elements
  $F_{x,y}(z,\om)\equiv F_{x,y}(0,\om)$ satisfy
$ F_{y+x,x}(0,\om) = F_{y,0}(0, T^x \om)$.
\end{enumerate}

Item (ii) corresponds to the eigenfunctions $\varphi_x$, which will be proved to be square summable, and even uniformly
exponentially decaying away from their individual "localization centers" $x\in\DZ^d$; (i) corresponds
to the eigenvalues $\lam_x$ associated with $\phi_x$. The last category (iii) covers the case of the
deterministic matrices (Hamiltonians) $H_\veps(\om)$ and various intermediate matrices used in the construction of approximate
eigenbases.

\paragraph{Local dependence and stochastic support.}

The starting point for the KAM procedure is an observation that the original, canonical
delta-basis in $\DZ^d$ is an approximate eigenbasis for the operator $H_\veps = \veps\Delta + V$,
with accuracy of order $\Ord{\veps}$, and the approximate AEVs $\lam^0_x = V(x;\om)$ exhibit local dependence
upon the values of the potential. This renders explicit the control of the small denominators
$\lam^0_x - \lam^0_y \equiv V(x;\cdot) - V(y;\cdot)$ appearing in the KAM procedure.

\bde
Let be given a measurable mapping $f$ from the space $\BM^{\rm diag}_{\DZ^d}$ of diagonal matrices $\Lam = \diag(\lam_x, x\in\DZ^d)$
to some measurable space $(\cA,\fB)$. The  stochastic support  of $f$, denoted $\Stoch{f}$, is the minimal
subset $\csS \subset\DZ^d$ such that $f$ is measurable with respect to the sub-sigma-algebra $\fF_\csS$
generated by the cylinder sets $\csL_{x,t} := \{\Lam\in\BM^{\rm diag}_{\DZ^d}:\, \lam_x \le t\}$, $x\in\csS$, $t\in\DR$.
\ede

The most important consequence of the finiteness (and uniform boundedness)  of the  stochastic support  of the AEVs
will be the following property: on each step $j\ge 0$, the differences $\lamj_x - \lamj_y$ (hence the
respective small denominators) for all $y\in \ball_{C L_j}(x)$ are invariant under the local transformations
$
V(x;\cdot) \mapsto V(x;\cdot) + t \one_{\ball_{C L_j}(x)}(x), \;\; t\in\DR.
$
This property provides an elementary tool for proving a satisfactory substitute for the Wegner bound.

\paragraph{Partitions and approximants of the potential.}
\label{sec:partitions}

The partitions $\cC_n$ introduced above form a monotone sequence: $\cC_{n+1}  \prec \cC_n$, i.e., each element of $\cC_n$ is a
union of some elements of the partition $\cC_{n+1}$.
Given $n\ge 0$, for each $\om\in\Om$ we denote by $\hk_n(\om)$ the unique index such that
$
\om \in C_{n,\hk_n(\om)}.
$
%
For each $N\ge 0$, introduce the approximant of   $v(\om,\th)$ given by \eqref{eq:def.v},
$$
v_N(\om,\th) = \sum_{0 \le n \le N} \;a_n\; \sum_{1 \le k \le K_n} \th_{n,k} \, \mchi_{n,k}(\om),
$$
the truncated potential $V_{N}$ and the truncated Hamiltonian $H^{(N)}_\veps$:
$$
V_{N}(x;\om,\th) := v_{N}(T^x\om,\th), \;\;\;\;
H^{(N)}_\veps := \veps \Delta + V_{N}.
$$
%
%
For any $N\ge 0$, denoting
$\| f \|_\infty := \sup\limits_{\om\in\Om} \|  f(\om, \cdot) \|_{L^\infty(\Th)}$,
we have
\bl
\label{eq:aN.decay.fast}
\norminf{ v - v_N } \le \sum_{n \ge N+1}  \eu^{-2 \monb n^2} \le \squart \eu^{-2 \monb } \eu^{-2 \monb N} a_N \,.
\el
It is important that the RHS is much smaller than the width ($a_N$)
of the distribution of random coefficients $a_N\th_{N,k}$,
$1 \le k \le K_N$ (recall: $\th_{N,k} \sim \Unif[0,1]$).
(This is why $a_n$ has to decay faster that exponentially.)
Let
\bl
\label{eq:def.tn.L}
\tn(L) = \lceil \ln^{2}(L) \rceil \,,
\el
then we have, for $L$ large enough,
\bl
\label{eq:Cn.sep.traj}
2^{-\tn(L)} < \shalf C_A^{-1}\, L^{-2A} .
\el
\UPA and \eqref{eq:Cn.sep.traj} guarantee that, for all $u\in\DZ^d$ and $\om\in\Om$, the phase points $\{T^x\om, x\in\ball_{L^2}(u)\}$ are separated
by the partition $\cC_{\tn(L)}$. (Replacing $L^2$ with $L^{bL}$, $b>0$, one has to replace $L^{-2A}$ with $L^{-bA}$ in \eqref{eq:Cn.sep.traj}.)

For further use, introduce the sigma-algebras $\fB_{N}^\msperp$ generated by the r.v.
\be
\label{eq:def.fB.n}
\{ \th_{n,k}, \, n\ne N, \,  k\in\llb 1,K_{n} \rrb \, \}\,, N\ge 0.
\ee
Conditional on $\fB^\msperp_{N}$, each value of the potential $V(\om,\th)$
with a fixed $\om\in\DT^\nu$ is an affine function of some $\th_{n,k}$ with $k = k(\om)$.
This significantly reduces the randomness in $V(\om,\th)$ but makes simpler and more transparent the analysis
of regularity of its probability measure.

\bre
\label{rem:Bxi.theta.hat}
Fix $L > 1$ and let $\tn_j = \tn(L_j)$, then by \eqref{eq:def.v}, for all $z\in\ball_{L^2}(0)$,
\bl
\notag
\tV_{\tn_j}(z; \om,\th)
&
= v_{\tn_j}(T^z\om, \th) = \Bxi_z(\th) + \hth_{j,z}(\th) \,,
\\
\Bxi_z(\th) &:= a_{\tn_j} \th_{\tn_j\,, \hk_{\tn_j}(T^z\om)} \,, \quad \hth_{j, z}(\th) := \sum_{n < \tn_j} a_n \th_{n,k} \mchi_{n,k}(T^z \om) \,.
\el
Here, all $\th \mapsto \hth_{j, z}$ are $\fB^\msperp_{\tn}$-measurable. The sets $\supp \mchi_{\tn,k}\subset\DZ^d$ with fixed $\tn$
are pairwise disjoint, thus by our hypothesis on the random variables $\th_{\bullet,\bullet}$,
the family random variables $\set{ \Bxi_z\,, \, z\in \ball_{L^{4}}(0) }$ on $\Th$
is independent, with a common probability distribution $\Unif([0, a_{\tn}])$.
Therefore, conditional on $\fB^\msperp_{\tn}$, the random variables $\th \mapsto V(z; \om,\th) = \Bxi_z + \hth_{j,z}$ are independent and uniformly distributed in individual
intervals of length $\ell = a_{\tn_j}$.
$\blacktriangleright$
\ere

Fix a length scale $L_j$, $j\ge 0$, let
$R_j = 2^{ -\tn_j}$, and partition $\Om=\DT^\nu$ into a union of $R_j^{-\nu }$ adjacent cubes $Q_{R_j}(\om_i)$,
$i\in \llb 1, R_j^{-\nu } \rrb$, of size $R_j$ and with centers $\om_i$ forming a periodic grid including $0\in\DT^\nu$.
Similarly, partition $\Om$ into adjacent cubes $Q_{r_j}(\tom_{i'})$ of size $r_j = L_j^{-6 A}\,$.


\bpr[Cf. {\lcite{C18b}{Corollary 2.1}}]
\label{cor:finite.cover.cubes}
For any $j\in \DN$, there exists a measurable partition of $\Om$,
$\cP_j = \big\{ \rP_{j,l}, \; 1\le l \le \cL_j \big\}$, $\cL_j \le 2^\nu r_j^{-\nu} = 2^{ \nu (\tn_j + 1)}$, such that,
for each $\rP_{j,l}\,$, the random vector $\Bfeta =(\eta_z\,, z\in\ball_{L_j}^2(0))$ given by
$$
\eta_z: \th \mapsto v_\tn\bpara{ T^z \om, \th ,  \, z\in\ball_{L_j}^2(0) }
$$
takes a constant value on $\rP_{j,l}\,$.
\epr

This result can be re-formulated as follows. Pick one point $\tau_{j,l}\in\rP_{j,l}$ per element $\rP_{j,l}$, and let $\cT_j$ be the entire (finite) collection of the
points $\tau_{j,l}$. Then, despite the fact that the argument $\om\in\Om$ takes an infinite number of values,
there exists only a finite number $\cL_j$ of possible random functions
\be
\notag
\ball_{L_j}^2(0) \ni z\mapsto v_\tn(T^z\om, \th)
= V_\tn(z; \om,\th)
= \sum_{\tau_{j,l}\in \cT_j } \one_{\rP_{j,l}}(\om) V_\tn(z; \tau_{j,l},\th) \,.
\ee

Proposition \ref{prop:prob.zeta.zeta} will allow us to avoid using a Wegner-type estimate.

\bpr[Cf. {\lcite{C18b}{Lemma 2.2}}]
\label{prop:prob.zeta.zeta}
Let be given $L_j$ and two functionals $\fa', \fa'': V \mapsto \DR$
of the functions $V:\, \DZ^d \to \DR$ satisfying for  some finite subsets $\Lam', \, \Lam''\subset \DZ^d$
\be
\notag
\all t\in\DR \quad
 \fa'[V + t\one_{\Lam'} ] = \fa'[V] + t, \quad
 \fa''[ V + t\one_{\Lam''} ] = \fa''[V] + t.
\ee
Assume that $\Lam' \cap \Lam'' = \varnothing$
and $\Lam' \cup \Lam''\subset \ball_{L_j^2}(0)$.
Let $V(x;\om,\th) = v(T^x\om, \th)$, $x\in\DZ^d$, and consider two r.v.
$\zeta'(\om, \th) = \fa'[V(\cdot; \om,\th) ]$,
$\zeta''(\om, \th) = \fa''[V(\cdot; \om,\th) ]$.
Then for some $C, C_1\in(0,+\infty)$ one has
\be
\notag
\prth{\th:\, \inf_{\om\in\Om} |\zeta'(\om,\th) - \zeta''(\om,\th) | \le s}
\le C_1 L_j^{C} a_{\tn_j}^{-1} \, s.
\ee

\epr

\section{KAM induction. Proof of Theorem \ref{thm:main.loc}}
\label{sec:KAM}

As in \cite{C18b}, we use the norms defined for the functions on $\DZ^d$
and for the matrices $A_{x,y}$ with entries indexed by $x,y\in\DZ^d$ (cf. \cite{BLS83}): with $m>0$ fixed, let
%
\bea
\label{eq:m-norm.f}
\mxnorm{ f } &= \sum_{y\in\DZ^d} \eu^{m|y-x|} \big| f(y)\big| , \;\; x\in\DZ^d,
\\
\notag
\vmnorm{ A } &= \sup_{x \in\DZ^d } \sum_{y \in\DZ^d} \eu^{m|y-x|} \big| A_{xy} \big| .
\eea
%
Note that $\norm{\cdot}_{0,x}$ is the conventional norm in $\ell^2(\DZ^d)$.
%
%
We also use a characteristic of finite-band matrices which we call \emph{the spread}
of a matrix $A$, denoted $\Spr{A}$:
\bl
\notag
\Spr{A} := \min\big[ r\in\DN:\; \all x,y\in\DZ^d \text{ with } |y-x|>r \;\;  A_{yx} = 0 \big].
\el
If $A$ is not a finite-band matrix, its spread is infinite, but we do not
encounter such situations. This meaning of the word "spread" is not traditional,
but we use it here occasionally, solely for the sake of terminological brevity.

Introduce an integer sequence (length scales) $\para{ L_j }_{j\in\DN}$ and decaying positive sequences $( \eps_j )_{j\in\DN}\,$,
$( \delta_j )_{j\in\DN}\,$, $( \beta_j )_{j\in\DN}\,$ of the form
\begin{align}
\label{eq:def.L.j}
L_j &= L_0 \, q^j \,, \quad
\,
\eps_j = (\eps_0)^{q^j}\,, \;\; q = 3/2, \quad \eps_0(\veps) := \veps^{1/4}  \,,
\\
\label{eq:def.delta.j.beta.j}
\delta_j &= a_{\tn_j} \beta_j \,,
\quad
\beta_j = \eu^{-\tn_j }\,, \quad \tn_j := \tn(L_j),
\end{align}
where $L \mapsto \tn(L) = \lceil \ln^{2}(L) \rceil$ (cf.~\eqref{eq:def.tn.L}).
Observe that
\bl
\notag
\delta_{j+1} /\delta_j
&
= \eu^{ -\para{\tn_{j+1}^2 - \tn_j^2} - \para{ \tn_{j+1} - \tn_j^{1} } }
%
\le \eu^{ - \tn_{j+1}^2/2 } \,.
\el

\vskip2mm
\noindent
\textbf{Inductive hypothesis} \KKj:
For all $0 \le i \le j$ there exists a measurable subset
$\tThi\subset\Th$ with $\DPTh\big\{ \tThi \big\} \ge 1 - \eps_i^{0+}$ such that for all $\th\in\Thj := \cap_{i=0}^j \tThi$  the following objects are well-defined:
\begin{enumerate}[\noindent\rm(K1)]
\setlength{\itemsep}{-1pt}
\setlength{\parskip}{-1pt}
  \item\label{lab:K1}
  $T$-covariant mappings from $\Om = \DT^\nu$ to $\Mat(\DZ^d, \DR)$ parameterized by $\th$:
%
\bl
\notag
(\om, \th) &\mapsto \Ui(\om,\th)\,, \quad \Ui_{yx}(\om,\th) =: \phii_x(y, \om, \th) = \phii_0(y-x;T^x\om,\th) \,,
\\
(\om, \th) &\mapsto \Psii(\om,\th)\,, \quad \Psii_{yx}(\om,\th) =: \psii_x(y, \om, \th) = \psii_0(y-x;T^x\om,\th) \,,
\\
\notag
(\om, \th) &\mapsto \Lami(\om,\th), \quad \Lami_{yx}(\om,\th) \; =: \delta_{yx} \lami_x(\om,\th) = \lami_0(T^x\om,\th)  \,,
\el
with $\norm{\ffii_x} = 1$ (here, $\norm{\cdot}$ is the usual norm in $\ell^2(\DZ^d)$).
We denote
\be
\label{eq:def.Fi}
\Fi(\om,\th) := (\Ui)^{-1} \Psii(\om, \th).
\ee
The following relations hold, for all $0\le i \le j$:
\vskip2mm
%
\item\label{lab:K2}
The matrix $\Ui(\om,\th)$ has the form
\be
\notag
\Uj = \one + \tDj \,,
\quad
\vmnorm{ \tDi(\om,\th) } \le \quart - \frac{1}{4^{i+2}} \,,
\ee
hence it is boundedly invertible by the Neumann series.
\vskip2mm

  \item\label{lab:K3} The matrices $\Ui, \Psii, \Lami$ satisfy the identity
\bl
\notag
H \Ui &= \Ui \Lami  +  \Psii  \,.
\el
%

  \item\label{lab:K4}
The discrepancy terms $\Psii_\bullet$ satisfy
\begin{align}
\label{eq:KLj.norm.psj}
 \sup_{\om\in\Om}  \;\; \sup_{x\in\DZ^d} \; \mxnorm{ \psii_x(\om, \th) } &\le  \eps_i \,,
\\
\label{eq:KLj.scalar.psij.x.phij.x}
\sup_{\om\in\Om} \;\; \sup_{x\in\DZ^d} \; \bigabs{ (\phii_x, \psii_x) }  &\le \eps_i^{\mmi{2}}.
\end{align}


  \item\label{lab:K5}
For $i=0$, one has $\lam^{0}_0(\om,\th) = v(\om,\th)$ and
\bl
\label{eq:i=0}
\phi^{0}_0(\om,\th) &= \one_{ \set{0} } \,, \;\;
\psi^{0}_0(\om,\th) &= \eps \sum_{y:\, |y| = 1} \one_{\set{0} } \,.
\el

For all $0 \le i \le j$, the objects $\lami_x,\phii_x,\psii_x$ are determined
by $\Lam^{(0)}$ and depend upon $(\om,\th)\in\Om\times\Th$  through the functions
$(\om,\th) \mapsto \lamz_x(\om,\th)$. Denoting by $\#[\Lam]$ the dependence of an object $\#$
upon $\Lam$, one has
%
\begin{align}
\label{eq:diag.cov.lam}
\all t\in\DR \qquad
\lami_x [ \Lamz + t\one_{\DZ^d}]  &= \lami_x[ \Lamz ]  + t,
\\
\notag
\phii_x [ \Lamz + t\one_{\DZ^d}]  &= \phii_x[ \Lamz ] ,
\\
\label{eq:diag.cov.psi}
\psii_x [ \Lamz + t\one_{\DZ^d}]  &= \psii_x[ \Lamz ]  .
\end{align}

  \item\label{lab:K6}
  The AEF $\phij_x$ have compact support, of size uniformly bounded in $x$:
\bl
\notag
\all\, x\in\DZ^d \quad \supp \, \phii_x \cup \supp \, \psi_x \subset \ball_{L_i}(x)  .
\el
%

  \item\label{lab:K7}
There exist $C>0$ and a function $\hj:\, \DZ^d  \times \DZ^d \to \DN$ such that
%
\begin{align}
\label{eq:KLj.spacings.j.4.Y.L}
\fs_i(x,y) &:= \inf_{\th\in \Thi} \; \inf_{\om\in\Om} \;\; \min_{\substack{x,y\in \ball( L^{2}_i, 0)\\ x \ne y}} \; | \lami_x(\om,\th) - \lami_y(\om,\th)| \ge 4\delta_i  \,,
\\[4pt]
\notag
\fs_i(x,y) & \ge \eu^{ - C \ln^4|x-y|} \; \prod_{i \ge \hj(x,y)} \para{ 1 - \eps_i } \gea \eu^{ - C \ln^4|x-y|} \,.
\end{align}
%

  \item\label{lab:K8}
The objects  $\lami_x, \phii_x, \psii_x$ have finite stochastic supports:
$$
\Stoch{\lami_x } \cup \Stoch{ \phii_x } \cup \Stoch{ \psii_x } \subset \ball_{\rrr{\Bc} L_i}(x).
$$

  \item\label{lab:K9}
For all $0 \le i \le j-1$, one has
%
\begin{align}
\notag
\sup_x | \lamione_x - \lami_x| \le \eps_i^{2-} \,,
\\
\notag
\sup_x \mxnorm{ \phiione_x - \phii_x } \le \eps_i^{1-} \,.
\end{align}
\end{enumerate}

\subsection{The base of induction}
\label{ssec:base.induction}

We start with the approximate eigenfunctions $\phiz_x(\om,\th)$ and eigenvalues $\lamz_x(\om,\th)$ as in \eqref{eq:i=0}.
It follows from the definition of the lattice Laplacian $\Delta$ that the discrepancies $\psiz_x$ in \eqref{eq:i=0} are correct:
\beal
\label{eq:check.psiz}
\psiz_x(\om, \th) = \big( \veps \Delta + V(x;\om,\th) - \lamz_x(\om,\th) \big) \one_x
=  \veps  \sum_{ |y-x|=1} \one_y \,.
\eeal

We assume that $\veps \le 1/(2d)^2$, and set
\bl
\notag
m(\veps) &:= \ln \veps^{-1/4} \tto{\veps \to 0} +\infty \,.
\el
Recall that we have introduced in \eqref{eq:def.L.j} the sequence $\eps_i = \eps_0^{q^i}$, $q = 3/2$, $i\ge 0$.

The relations \eqref{eq:diag.cov.lam}--\eqref{eq:diag.cov.psi} with $i=0$ follow from
the explicit form of $\lamz_x(\om,\th) = v(T^x\om, \th)$ and from the $(\om,\th)$-independence of $\phiz_x$ and $\psiz_w$.
Also, $(\psiz_x, \phiz_x) \equiv 0$, which is stronger than (K\ref{lab:K8}) with $j=0$.

We often use notations like "$\eps_i^{\mpmi{b}}\,$" as shortcuts for
\emph{"$\eps_i^{b\pm c}$ with $c>0$ that can be chosen (before the induction starts) as small as necessary"}.
Among such implicit exponents, the one figuring in \eqref{eq:KLj.spacings.j.4.Y.L} is quite important, so we denote it by $\ups$ and specify
its relations to other key quantities.

Recalling $\tn(L) = \ln^2(L)$, denote for brevity $\tn = \tn(L_0)$, and
assume that
\bl
\notag
\veps \le \eu^{ - 8 \monb \ups^{-1} \tn^2} = \eu^{ - 8 \monb \ups^{-1} \ln^4(L_0)} \,.
\el
Then
\bl
\notag
\delta_0(\veps)
&
= \eu^{ - \monb \tn^2(L_0) - \monb \tn(L_0)} \ge  \frac{1}{5} \eu^{ - 2 \monb  \tn^2}  \ge \frac{1}{5} \veps^{\ups/4} = \frac{1}{5} \eps_0^{\ups}(\veps) \,.
\el

Now we turn to the norm-estimates of the discrepancies $\psiz_\bullet$ (cf. \eqref{eq:KLj.norm.psj} in (K\ref{lab:K4})).
By covariance, it suffices to check \eqref{eq:m-norm.f} with $x=0$: by \eqref{eq:check.psiz}, one has
$$
\| \psiz_0 \|_{m,0}
= \veps\sum_{y:\, |y|=1} \eu^{m|y|}1 = 2d \eu^{m} \, \veps
= 2d \veps^{1/2} \cdot \eu^{m} \veps^{1/2} \le 1 \cdot \eps^{-\frac{1}{4} + \half}
= \eps^{1/4} \,.
$$
Taking $\veps>0$ in \eqref{eq:def.H} small enough, one can have both $\eps_0>0$ arbitrary small
and the $m$-norm estimate $\eqref{eq:KLj.norm.psj}$ holding with $m>0$ arbitrarily large.

Fix $L_0>1$, denote $\tn = \tn(L_0)$, and consider the approximate eigenvalues
\bl
\notag
\lamz_x(\om,\th) = v(T^x\om\,, \th) \,, \quad |x| \le L_0^{2} \,.
\el
Conditional on $\fB_{\tn}^\msperp$ (cf. \eqref{eq:def.fB.n}), the AEV $\lamz_x(\om,\th)$ is a measurable function of
$\{ \th_{\tn}\,, \, 1\le k \le 2^{\tn} \}$, and letting $k(x)$ be such that $T^x\om \in \supp \mchi_{\tn, k(x)}\,$, we get
$$
\lamz_x(\om,\th) = \th_{\tn, k(x)} c_x(\om,\th) + c_x(\om,\th)  \,,
$$
where $\th \mapsto c_x(\om,\th)$ is $\fB_{\tn}^\msperp$-measurable.
The cubes $C_{\tn,k}$ (cf. \eqref{eq:def.C.n.k}) separate the points $\set{T^x\om\,, |x|\le L_0^{2}}$ (cf. \eqref{eq:Cn.sep.traj}),
so we can assess, for a fixed $\om\in\Om$,
the probability of the event $\cE_{x} = \set{\th:\, \abs{\lamz_0(\om,\th) - \lamz_x(\om,\th)} < 4 \delta_0 }$.
Consider the partition $\cP_j = \set{\rP_{j,l}}$ from Proposition \ref{cor:finite.cover.cubes}, then for any $\tau_{j,l}\in\rP_{j,l}$

\bl
\label{eq:v.tn.-tn}
&{\max_{ \substack{x,y\in \ball\\x\ne y}}} \;
\prth{ \abs{ v_{\tn_j}(T^x\om,\th) - v_{\tn_j}(T^y\om,\th) } \le s \cond \fB_{\tn}^\msperp }
\\
&\quad
\le 2 \frac{|\ball|^2}{2}  \prth{ \abs{\th_{\tn, k(x)} - \th_{\tn, k(y)} + c'_{x,y}} \le s \cond \fB_{\tn}^\msperp }
\le  L_0^{ 4d } a_\tn^{-1} s
\el
with some $\fB_{\tn}^\msperp$-measurable $c'_{x,y}=c'_{x,y}(\om,\th)$.
Let $s = 5 \beta_0 a_\tn$, $\beta_0 = \eu^{- \tn }$, $\delta_0 = \beta_0 a_{\tn} = \eu^{ - \tn^2 -  \tn}$,
so $\beta_0 = \delta_0^{ \mpi{0}}$.
Taking expectation in \eqref{eq:v.tn.-tn}, we get
\bl
\notag
\prth{ \abs{\th_{\tn, k(x)} - \th_{\tn, k(y)} + c'_{x,y} }\le 5 \beta a_\tn }
&
\le \eu^{  -2\monb \ln^4(L_0) +  C\ln^2 L_0}
\le \eu^{  -\monb \ln^4(L_0) }
\el
provided
%
$L_0$ is large enough.
Now, let
\bl
\notag
\Th^{0}(\om) := \set{ \inf_\om \; \abs{\lamz_0(\om,\th) - \lamz_x(\om,\th)} \ge 4 \delta_0 }.
\el
Then making use of constance of $\mchi_{\tn,k}$ on any $\rP_{0,l}$, the bound
$\card \cP_0
\le 2^{\nu (\tn + 1)}$ (cf. Proposition \ref{cor:finite.cover.cubes}),
and the approximation bound \eqref{eq:aN.decay.fast}, we conclude:
%
\bl
\notag
\inf_{\th \in \Th^0}\; \inf_\om \; \abs{\lamz_0(\om,\th) - \lamz_x(\om,\th)} \ge  4 \delta_0 \,,
\\
1 - \prth{\Th^0 }
\le \eps_0^{\mpi{0}} \,.
%
\el

\subsection{The inductive step}

Below we use sometimes for brevity the notation like $a(j) \lea b(j)$ for quantities
dependent upon the scale $L_j$, meaning that $a(j) \le C b(j)$ for some finite constant $C$ and all $j\ge 0$.
The subscript $\veps$ in $H_\veps$ will be often omitted.

\btm
\label{thm:LAM.induction}
For any $j\ge 0$, \KKj implies \KKone.
\etm

\proof Fix $j\ge 0$ and assume \KKj.

\par\vskip1mm\noindent
\textbf{Step 1. The Gram matrix.}
Let us show that the Gram matrix $\Cj = (\Uj)^{\top} \Uj$ of the Riesz basis $\{\phij_\bullet\}$ is close to $\one$, viz.
$
\Cj = \one + \Dj, \; \; \| \Dj \|_m = \Ord{\eps^{1-}_j} .
$
It will imply the convergence of Neumann's series for $\big(\one + \Dj \big)^{-1}$, so
$$
(\Cj)^{-1} = (\one + \Dj)^{-1} = \one - \Dj + \Ord{\| \Dj \|_m^2} = \one + \Ord{\eps^{1-}_j}.
$$
By symmetry of $H$, we have
\begin{align}
\label{eq:C.j.xy.phi.psi}
(\lamj_y -  \lamj_x) \Cj_{yx}
&
= (\phij_y\,, \psij_x) - (\psij_y \,, \phij_x) \,,
\\
\notag
|\lamj_y - \lamj_x| \cdot \abs{ \Cj_{yx} }
& \le \abs{ ( \phij_y,  \psij_x) } + \abs{ ( \psij_y, \phij_x) } \,.
\end{align}
For all $x \ne y$, there are two alternatives:
\par\noindent
$\bullet$  $|x - y| > 2 L_j$, so $\supp \psij_x \cap \supp \phij_y = \supp \psij_y \cap \supp \phij_x = \vide$,
then
\begin{align}
\notag
(\psij_y, \phij_x) &= (\phij_y, \psij_x) = 0 \,,
\quad
\Cj_{yx} = 0 \,.
\end{align}
\par\noindent
$\bullet$ $|x - y| \le  2 L_j$, then $|\lamj_x - \lamj_y| \ge 4 \delta_j$ by (K\ref{lab:K7}), so
\be
\notag
| \Cj_{yx} | = |( \phij_y,  \phij_x)|
\le \frac{\big|( \phij_y,  \psij_x)\big| + \big|( \psij_y, \phij_x)\big| }{ 4 \delta_j } \,.
\ee
(K\ref{lab:K9}) and \eqref{eq:i=0} imply that $\mxnorm{ \phij_x } \le 1 + \sum_{i} \eps_i^{1-} \le 2$,
while $\mxnorm{\psij_x} \le \eps_j^{\onemi}$ by \eqref{eq:KLj.norm.psj},
hence
\be
\label{eq:scalar.phi.small}
\bigabs{ ( \psij_y,  \phij_x) } + \bigabs{ ( \phij_x,  \psij_y) } \lea L_j^{d-1} \eu^{-m|x-y|} \delta_j^{-1} \eps_j
\lea \eu^{-m|x-y|}\, \eps_j^{1-} \,.
\ee
Recalling that $\Cj_{xx} = \| \phij_x \|^2=1$ by (K\ref{lab:K1}), we get
\bl
\label{eq:Cxy.bound}
\Cj_{yx}
\left\{
  \begin{array}{ll}
    = 1 , & \hbox{if $x=y$;}
\\
\equiv \Dj_{yx} \,, \;\;   \abs{ \Dj_{yx} }  \lea  \eu^{-m|x-y|} \, \eps^{1-}_j , & \hbox{if $0 < |x-y| \le R_j$;}
\\
    =0 , & \hbox{if $|x-y| > R_j$.}
  \end{array}
\right.
\el
Thus $\Cj = \one + \Dj \,$, where
\begin{align}
\label{eq:norm.Dj}
\vmnorm{  \Dj }
&
= \sum_{z\ne 0: \, \Cj_{0 z}\ne 0} \eu^{m|z|} \, \bigabs{ \Cj_{0z} }
\lea \underbrace{\eps_j^{1-}}_{\text{by \eqref{eq:Cxy.bound}} } \sum_{|z|\le  2 L_j }  \, 1 \lea L_j^{d} \, \eps_j \le \eps_j^{1-} ,
\end{align}
and so it follows from $\big( (\Uj)^\top  - (\Uj)^{-1} \big) \Uj = \Dj$ and $\vmnorm{  (\Uj)^{-1} } \le 2$ that
\begin{align}
\label{eq:UT.minus.Uinv.exact}
(\Uj)^\top  - (\Uj)^{-1}  &=  \Dj \, \Ujinv \,,
\\
\label{eq:UT.minus.Uinv.norm}
\vmnorm{ (\Uj)^\top  - (\Uj)^{-1} } & \le 2 \vmnorm{ \Dj } \le 4 \eps_j^{1-} \,.
\end{align}
By (K\ref{lab:K3}), the matrix $\Uj = \one + \tDj$ is invertible by Neumann series, and
\bl
\notag
\vmnorm{ (\Uj)^{-1} -\one }  & \le \sum_{k\ge 1} \vmnorm{ \tDj }^k \le \sum_{k\ge 1} \left( \quart - \frac{1}{4^{j+2}} \right)^k
\le \frac{1}{3} - \frac{1}{3 \cdot 4^{j+1} } \,,
\el
whence
\bl
\notag
\max_{0 \le i \le j} \; \max\Bbrak{ \, \vmnorm{ \Uj }\,,  \vmnorm{ (\Uj)^{\top} } \,, \, \vmnorm{ (\Uj)^{-1} } \, }
&
< 2.
\el
%

\par\vskip3mm\noindent
\textbf{Step 3. Construction of the new AEF.}
Introduce a  matrix
\bl
\label{eq:def.tQj}
\tQj := (\Uj)^\top \Psij \,, \quad \tQj_{yx} = (\psij_x\,, \phij_y)\,,
\el
as an approximant to $\Fj = \Ujinv \Psij$ with the $\vmnorm{\!\cdot\!}$-accuracy of $\eps_j^{\onemi}$,
owing to \eqref{eq:UT.minus.Uinv.exact}--\eqref{eq:UT.minus.Uinv.norm},
and its truncated variant $\Qj$:
%
\bl
\label{eq:def.Qj}
\Qj_{yx} := \tQj_{yx} \one_{ \set{ |x-y| \le c L_j} },
\el
with $c \in (0,1)$ to be specified later.
Observe that
\bl
\notag
\all C,c >0 \;\; \all m  \ge \frac{ C \ln \eps_0^{-1}}{  c L_0 } \;\;
\quad\;
\eu^{ - m  c L_i }
&
= \eu^{ - m  c L_0 q^i }
\le \bpara{ \eps_0^{ q^{i} } }^{ C }
= \eps_{i}^{ C } \,.
\el
\noindent
$\Qj$ has a small norm and finite spread:
\bl
\label{eq:mnorm.Spr.Qj}
\vmnorm{ \Qj }
&
\le \vmnorm{ \tQj } \le \vmnorm{ \UjT } \cdot \vmnorm{ \Psij } \lea \eps_j \,,
\\[3pt]
\Spr{\Qj} & \le c L_{j+1} \,.
\el
Next, define a matrix  $\Mjo$ by its entries:
\be
\label{eq:def.M.j+1}
\Mjo_{yx} =
\begin{cases}
\frac{ \Qj_{yx} }{ \lamj_x - \lamj_y }, & \text{ if  $y \ne x$ and $\Qj_{yx} \ne 0$,}
\\
0,  & \text{ otherwise}.
\end{cases}
\ee
All the entries $\Mjo_{yx}$ are indeed well-defined, thanks to the hypothesis (K\ref{lab:K7}) (cf. \eqref{eq:KLj.spacings.j.4.Y.L}).
$\Mjo$ defines an operator in the space of compactly supported functions on $\DZ^d$,
on which one has (cf. \eqref{eq:KLj.spacings.j.4.Y.L} and \eqref{eq:mnorm.Spr.Qj})
\bl
\label{eq:norm.Mjo}
\vmnorm{ \Mjo } & \le \eps_j \, \delta_j^{-1} \le \eps^{1-}_j \,,
\\[2pt]
\Spr{ \Mjo } & = \Spr{ \Qj } \le c L_{j+1} \,,
\el
so $\Mjo$ defines also a bounded operator in $\ell^2(\DZ^d)$. The column-vectors of the matrix
\be
\notag
\tUjo := \Uj\big(\one + \Mjo\big)
\ee
form a Riesz basis in $\ell^2(\DZ^d)$, because both $\Uj$ and $\one + \Mjo$ are boundedly invertible. Denoting these column-vectors by $\tffijo_\bullet$, we have
\be
\label{eq:def.phijo}
\tffijo_x = \phij_x + \sum_{z\ne x} \Mjo_{zx} \phij_z \,, \quad x\in\DZ^d \,.
\ee
By induction, $\phij_\bullet$ and $\Mjo_{\bullet}$ are invariant under the flow $\Lamz \mapsto \Lamz + t\one_{\DZ^d}$,
and so are, therefore, $\tffijo_\bullet$ which are functions of $\phij_\bullet$ and $\Mjo_{\bullet}$.
The normalization of $\tffijo_\bullet$ will be performed at \textbf{Step 7}.
By the inequalities $\Spr{AB} \le \Spr{A} + \Spr{B}$ and $\Spr{A+B} \le \max\brak{ \Spr{A}\,, \Spr{B} }$, one has, for $c>0$ small enough and $L_0$ large enough:
\begin{align}
\notag
\Spr{ \tUjo } \le \Spr{ \Uj } + \Spr{ \Mjo } &\le 2c L_j + 1 \le 3c L_j \,,
\\
%
\notag
\supp \tffijo_x \bigcup \Stoch{\tffijo_x} &\subset \ball_{ 3c L_j }(x) \,.
\end{align}

\noindent
By expansion in the Neumann series, convergent thanks to \eqref{eq:norm.Mjo}, we have
\begin{align}
\label{eq:inverse.1.plus.Mjo}
(\one + \Mjo)^{-1}
&
= \one - \Mjo + (\Mjo)^2 - (\Mjo)^3(1 + \Mjo)^{-1} \,,
\end{align}
so the inverse $(\one + \Mjo)^{-1}$ can be replaced by $\one - \Mjo + (\Mjo)^2$
with accuracy $\bigOrd{\, \vmnorm{ \Mjo }^3}$; the explicit formula \eqref{eq:inverse.1.plus.Mjo}
will be used later.

\par\vskip2mm\noindent
\textbf{Step 4. Action of $H_\veps$ on $\tffijo_\bullet$.}
By definition of $\tQj$ and $\Dj$, we have
\bl
\notag
\Ad_{\Uj} [ H ]
&
= \Lamj + \tQj  - \Dj \Fj \,.
\el
By straightforward calculations making use of the identities $\brak{ \Lamj\,, \Mjo} = -\Qj$ (cf. \eqref{eq:def.M.j+1}) and \eqref{eq:inverse.1.plus.Mjo},
we obtain the representation
\bl
\notag
\Ad_{\tUjo} [ H ]
&
=
(\one + \Mjo)^{-1} (\Uj)^{-1} H \Uj (\one + \Mjo)
= \Lamj + \Wjo + \Zjo \,,
\el
%
%
\noindent
where
\begin{align}
\label{eq:W.j+1}
\Wjo &=  \brak{ \Qj\,, \Mjo } + \Dj \Fj  +  (\Mjo)^2 \Lamj  - \Mjo \Lamj \Mjo  \,,
\\
\label{eq:Z.j+1}
\Zjo
&
=
- \Mjo \Qj \Mjo +  \brak{ \Dj \Fj\,, \Mjo } - \Mjo \Dj \Fj \Mjo
\\
\notag
&\quad
+ (\Mjo)^2 \Lamj \Mjo  - (\Mjo)^2 \Dj \Fj (\one + \Mjo)
\\
\notag
&\quad
- (\Mjo)^3 (\one + \Mjo)^{-1}\, \para{ \Lamj + \tQj  - \Dj \Fj} \, (\one + \Mjo).
\end{align}
Equivalently,
\bl
\label{eq.Ad.U.j+1.H}
(\tUjo)^{-1}  H \,\tUjo
&= \Lamjo + \Fjo \,,
\el
where $\Lamjo$ and $\Fjo$ are defined by their matrix elements:
\begin{align}
\label{eq:def.Lam.j+1.F.j+1}
\Lamjo_{yx} &= \Lamj_{yx} + \delta_{yx} \, \Wjo_{xx} \,,
\\
\label{eq:def.F.j+1}
\Fjo_{yx} &= (1-\delta_{yx}) \,\Wjo_{yx} + \Zjo_{yx}.
\end{align}
Define the new AEV $\lamjo_\bullet$:
\bl
\label{eq:def.lam.j+1}
\lamjo_x := \Lamjo_{xx} = \lamj_x + \Wjo_{xx} \,,
\el
then, by virtue of \eqref{eq.Ad.U.j+1.H}, we have:
\bl
\label{eq:lam.j+1.-.lam.j}
\sup_x\; \abs{ \lamjo_x - \lamj_x} \le \sup_x\; \abs{ \Wjo_{xx} } \le \eps_j^{\mmi{2}} \,.
\el
An equivalent form of \eqref{eq.Ad.U.j+1.H} is
\bl
\label{eq:H.U.j+1.Psi.j+1}
H \tUjo &= \tUjo \Lamjo + \Psijo \,,
\quad
\Psijo = \tUjo \Fjo \,,
\el
and since $\vmnorm{\tUjo} \le 2$, one has:
\bl
\notag
\vmnorm{ \tUjo \Zjo } &\lea \vmnorm{\Fjo } \vmnorm{ \Mjo }^2 + \vmnorm{\Fjo} \vmnorm{\Dj} \vmnorm{\Mjo} + \vmnorm{ \Mjo }^3 .
\el
Since $\Psijo = H \tUjo - \tUjo \Lamjo$, where $\Spr{H} = 1$, $\Spr{\Lamjo} = 0$, we have
\bl
\notag
\Spr{ \Psijo } &
%
\le \Spr{\tUjo } + 1.
\el
Finally, note that $\Wjo$ is invariant under $\Lamz \mapsto \Lamz + t\one_{\DZ^d}$ (cf. \eqref{eq:W.j+1}),
thus $\lamjo_\bullet\brak{\Lamz + t\one_{\DZ^d}} = \lamjo_\bullet\brak{\Lamz} + t$, just like $\lamj_\bullet$ (cf. \eqref{eq:diag.cov.lam}).

\par\vskip2mm\noindent
\textbf{Step 5. Norm of the discrepancy.}
Using the bounds
$\vmnorm{ \Fjo } \le \vmnorm{ \Wjo } + \vmnorm{ \Zjo }$,
$\vmnorm{ \tUjo } \le 2$, $\vmnorm{ \Qj } \le \eps_j$ (cf.~\eqref{eq:mnorm.Spr.Qj}),
$\vmnorm{ \Mjo } \le \eps_j^{\onemi}$ (cf. \eqref{eq:norm.Mjo}),
$\vmnorm{\Dj}\le \epsj^{\onemi}$ (cf. \eqref{eq:norm.Dj}),
$\vmnorm{\Fj}\le \epsj^{\onemi}$ (cf. \eqref{eq:def.Fi} and \eqref{eq:KLj.norm.psj}),
we get
\begin{align}
\label{eq:norm.W.j+1}
\vmnorm{ \Wjo }
&
\lea \vmnorm{ \Qj } \, \vmnorm{ \Mjo } + \vmnorm{ \Lamj } \, \vmnorm{ \Mjo }^2 + \vmnorm{ \Dj } \, \vmnorm{ \Fj } \le \eps_j^{\mmi{2} } ,
\\
\label{eq:norm.Z.j+1}
\vmnorm{ \Zjo }
&
\lea  \vmnorm{ \Dj } \, \vmnorm{ \Fj } \,  \vmnorm{ \Mjo } + \vmnorm{ \Mjo }^2  \vmnorm{ \Fj } + \vmnorm{ \Mjo }^3  \le \eps_j^{\mmi{3}}  \,.
\end{align}
Further, on account of $\Fjo_{yx} = (1 - \delta_{yx}) \Wjo_{yx} + \Zjo_{yx}$ (cf. \eqref{eq:def.F.j+1}),
%
\bl
\notag
\vmnorm{ \Fjo }  &\le \eps_j^{\twomi } \,, \quad
\vmnorm{ \Psijo }  = \vmnorm{ \tUjo \, \Fjo} \le \eps_j^{\twomi } .
\el

\par\vskip2mm\noindent
\textbf{Step 6. Perturbations of the AEF.}
By $\tUjo = \Uj(\one + \Mjo)$, we have
%
\bl
\notag
\vmnorm{ \tUjo - \Uj } &\le \vmnorm{ \Uj } \vmnorm{ \Mjo } \le 2 \vmnorm{ \Mjo } \le \eps_j^{ \onemi } \,,
\\
\mxnorm{ \tffijo_x - \phij_x } &\le  \eps_j ^{ \onemi}.
\el
Now, we prepare for the proof of the assertion \eqref{eq:KLj.scalar.psij.x.phij.x} with $i=j+1$
(to be completed on Step \textbf{7}).
By definition of $\psijo_\bullet$ and $\Fjo$ (cf. \eqref{eq:H.U.j+1.Psi.j+1},\eqref{eq:def.Lam.j+1.F.j+1}),
\bl
\label{eq:def.psi.j+1}
\psijo_x = \sum_{z} \bpara{ (1 - \delta_{zx} \Wjo_{yx}) + \Zjo_{yx} } \, \tffijo_z
\el
By the norm estimate \eqref{eq:norm.Z.j+1}, we have, on account of $\eps_{j+1} = \eps_j^q = \eps_j^{3/2}$,
\bl
\label{eq:scalar.psijo.x.tffijo.x.Z}
\Bigabs{ \sum_{z} \Zjo_{zx}  \para{ \tffijo_z \,, \, \tffijo_x } } \lea \vmnorm{ \Zjo } \le \eps_{j}^{ \mmi{3} } \le \eps_{j+1}^{ \mmi{2} } \,.
\el
Further, since $\mxnorm{ \tffijo_y - \phij_y}\le \eps_j ^{ \onemi}$ for all $y\in\DZ^d$,
and $\abs{ (\phij_x \,, \phij_z ) } \le \eps_j^{\onemi}$  for $z\ne x$ by \eqref{eq:scalar.phi.small},
we also have
$\abs{ (\tffijo_x \,, \tffijo_z ) } \le C \eps_j^{ \onemi} \le \eps_j^{ \onemi}$,
yielding
\bl
\label{eq:scalar.psijo.x.tffijo.x.W}
\Bigabs{ \sum_{z} (1 - \delta_{zx}) \Wjo_{zx}  \para{ \tffijo_z \,, \, \tffijo_x } } \le (C L_j)^d \vmnorm{ \Wjo } \,  \eps_j^{ \onemi} \le \eps_{j+1}^{ \mmi{2} } \,.
\el
Collecting \eqref{eq:scalar.psijo.x.tffijo.x.Z}--\eqref{eq:scalar.psijo.x.tffijo.x.W}, we get an analog of \eqref{eq:KLj.scalar.psij.x.phij.x}: \,
$\abs{ \para{ \psijo_x\,, \, \tffijo_x } }  \le \eps_{j+1}^{ \twomi }$.

\par\vskip2mm\noindent
\textbf{Step 7. \textbf{Normalization of the AEF.}}
Introduce the  normalized AEF
\bl
\label{eq:def.phi.j+1}
\phijo_x := \| \tffijo_x \|^{-1}\, \tffijo_x
\el
($\norm{\cdot}$ stands for the $\ell^2(\DZ^d)$-norm). Thus $\Stoch{ \phijo_x }  = \Stoch{ \tffijo_x } \subset \ball_{L_{j+1}}(x)$.
Since $\norm{\phij_\bullet}=1$, we have
\bl
\notag
\norm{\tffijo_x}^2 - 1 = 2 (\tffijo_x - \phij_x\,, \phij_x) + \norm{\tffijo_x - \phij_x}^2 \,.
\el
Recalling $(1 - \delta_{zx}) \Cj_{zx} = \Ord{ \eps_j^{\mmi{1}}}$,
it follows from \eqref{eq:def.phijo} and \eqref{eq:Cxy.bound} that
\bl
\notag
\abs{ (\tffijo_x - \phij_x\,, \phij_x) }
&
\le \sum_{z \ne x} \abs{ \Mjo_{zx} } \, \abs{ \Cj_{zx} }
\lea L_j^d  \eps_j^{\onemi} \cdot \eps_j^{\onemi} \le \eps_j^{\twomi} ,
\el
and $\norm{\tffijo_x - \phij_x} \le \eps_j^{\onemi}$ (cf. \textbf{Step 6}), thus
$\bigabs{ \norm{\tffijo_x} - 1 } \le \eps_j^{\twomi} \,$,
and \eqref{eq:KLj.scalar.psij.x.phij.x} with $i=j+1$ follows from
\begin{align}
\notag
\mxnorm{ \phijo_x - \tffijo_x } &\le \eps_{j}^{ \twomi } \,,
\\[2pt]
\label{eq:norm.phijo-phij}
\mxnorm{ \phijo_x - \phij_x } &\le \mxnorm{ \phijo_x - \tffijo_x } + \mxnorm{ \tffijo_x - \phij_x } \le \eps_{j}^{ \onemi } \,,
\\[2pt]
\notag
\left( \psijo_x, \, \phijo_x \right) &\le \eps_{j+1}^{\mmi{\redtwo}} \,.
\end{align}

\vskip2mm
\par\vskip2mm\noindent
\textbf{Step 8.} The assertion (K\ref{lab:K2}) is proved as in \lcite{C18b}{Section III, Step 8}.

\vskip2mm
\noindent
\textbf{Step 9. The local spectral spacings.}
It follows from the explicit formulae \eqref{eq:def.M.j+1}, \eqref{eq:def.phijo}, \eqref{eq:W.j+1}--\eqref{eq:Z.j+1}, \eqref{eq:def.Lam.j+1.F.j+1}, \eqref{eq:def.phi.j+1},
along with the linear growth rate bound on the diameters of the stochastic supports of the AEF/AEV, that the local deformations of the potential of the form
\be\label{eq:step9.V.t}
V(x; \om, \th) \rightsquigarrow V(x;\om,\th) + t\one_{\ball_{L_j^2}}(x), \;\; t\in\DR\,,
\ee
leave invariant the AEFs $\ffijo_y$ and the discrepancies $\psijo_y$ with
$|y-x|\lea L_j^2$, while the AEVs undergo a common shift $\lamjo_y \rightsquigarrow \lamjo_y + t$ under \eqref{eq:step9.V.t}.

\bre
In fact, $L_j^2$ could have been replaced with $CL_j$, with an appropriate $C>0$; using $L_j^2$ merely makes it more clear that we eliminate
the small denominators $\abs{\lami_x - \lami_y}$, by parameter exclusion in $\Th$, well before they can ever become dangerous in the inductive construction via $\Mi_{yx}$.
\ere

The strategy of the proof of (K\ref{lab:K7}) is as follows.
The required $\Th$-probability estimate is inferred from Proposition \ref{prop:prob.zeta.zeta}. Since the latter operating with independent random variables,
we condition on $\fB^\msperp_{\tn_{j+1}}$, $\tn_{j+1}:= \tn(L_{j+1}$, thus rendering samples in the $L_{j+1}^2$-balls independent.

Fix $x,y\in\DZ^d$ with $|x-y|\le L_{j+1}^2\,$, and consider two alternatives.

\vskip2mm
\noindent
\textbf{(I)} $L_{j}^2 < |x-y|\le L_{j+1}^2$.

In this case, we define the moment $\hj(x,y) := j+1$ when a lower bound on $\abs{\lam^{\bullet}_x - \lam^{\bullet}_y}$ is established for the first time,
at the price of exclusion of some subset of $\Th$. Before this moment, i.e., for $0 \le i < \hj(x,y)$, we have no effective control of $\abs{\lami_x - \lami_y}$:
the latter may be abnormally small. However, $\abs{\lami_x - \lami_y}^{-1}$ never appears in the inductive procedure until the moment $\hj(x,y)$.

To prove the required bound, we argue as in \cite{C18b};
the only technical distinction is that we define here $\tn(L) \sim \ln^2(L)$ instead of $\tn(L) \sim C\ln L$. (Actually, the threshold $\tn(L)$ can be chosen in various ways,
including $\tn(L) = L^c$ with a suitable (sufficiently small) $c>0$.)
Specifically, let
\bl
\notag
\tTh^{j+1} = \big\{\th\in\Th:\, \inf_{\om\in\Om} \, \abs{ \lamjo_x(\om,\th) - \lamjo_y(\om,\th) } \ge 5 \delta_{j+1} \big\} .
\el
By the inductive hypothesis  (K\ref{lab:K5}) (cf. \eqref{eq:diag.cov.lam}), we can apply Proposition \ref{prop:prob.zeta.zeta}
with $\fa'[V] = \lamjo_x$, $\Lam' = \ball_{L^2_{j+1}}(x)$, $\fa''[V] = \lamjo_y$, $\Lam'' = \ball_{L^2_{j+1}}(y)$, and obtain:
\bl
\notag
1 - \prth{ \tTh^{j+1} }
&
\le C' L_{j+1}^C a_{\tn_j}^{-1} \delta_{j+1}
%
%
\le  \eu^{ C \ln L_j - \tn_{j+1} }
\le  \eu^{ - \frac{1}{2}  \ln^2 L_{j+1} }
\el
(cf. \eqref{eq:def.delta.j.beta.j}). Furthermore, $\norminf{V - \tV_{\tn} } \le \eu^{- 2 \tn} a_\tn  = \ord{ \beta_{j+1} a_\tn} = \ord{\delta_{j+1}}$
(cf. \eqref{eq:aN.decay.fast} and \eqref{eq:def.delta.j.beta.j}),
thus $| \lamjo_z - \tlamjo_z | = \ord{\delta_{j+1}}$ for all $z$, and so
\bl
\notag
\all \th \in \tTh^{j+1}(\om) \quad \abs{ \lamjo_x(\om,\th) - \lamjo_y(\om,\th) } \ge  (5 - \ord{1}) \delta_{j+1} \,.
\el
It follows from $|x-y| \ge  L_{j+1}^2$ that $\tn \le C \ln^2|x-y|$, hence
\bl
\label{eq:spacing.when.born}
\all \th \in \tTh^{j+1}(\om) \quad \abs{ \lamjo_x(\om,\th) - \lamjo_y(\om,\th) } \ge \eu^{-C' \ln^4 |x-y|}.
\el
As we shall see in the case \textbf{(II)} below, the lower bound \eqref{eq:spacing.when.born}, with $x$ and $y$ fixed, is essentially preserved
on all subsequent induction steps.

\noindent
\textbf{(II)} $|x-y|\le L_{j}^2$.

In this case, $|\lam^\bullet_x - \lam^\bullet_y|$ has been first assessed on the step $\hj(x,y) \le j$:
\bl
\notag
\abs{ \lam^{\hj(x,y)}_x - \lam^{\hj(x,y)}_y } \ge (5g - \ord{1}) \delta_{\hj(x,y)} \,.
\el
By the perturbation bound $\abs{ \lam^{i+1}_\bullet - \lam^{i}_\bullet} \le \eps_j^{\mmi{2}}$ (cf. \eqref{eq:lam.j+1.-.lam.j}), we thus have
\bl
\notag
&{ |\lamjo_x - \lamjo_y|} \ge (5g - \ord{1}) \delta_{\hj(x,y)}   - \sum_{\hj(x,y) \le i \le j} 2\sup_z \; \abs{ \lam^{i+1}_z - \lami_z }
\\
&
\ge \para{ 5g - \ord{1} } \para{ 1- \delta_{j}^{-1} \eps_j^2} \delta_{\hj(x,y)}
\ge (5g - \ord{1})\para{1 - \eps_j} \delta_{\hj(x,y)}
\el
where $\ord{1}$ is relative to $\veps\to 0$.

\vskip2mm
\noindent
\textbf{Step 10. Supports and stochastic supports.}


First, note that $\supp \phijo_x = \supp \tffijo_x$, $\tffijo_x - \phij_x = \sum_{y\ne x} \Mjo_{yx} \phij_y$ (cf. \eqref{eq:def.phijo}),
$\Spr{\Mjo} \le \Spr{\Qj} \le cL_j$, and $\diam \supp \phij_y \le L_j$ for any $y$, thus
$\diam \supp \tffijo_x  \le (1 + c)L_j) < L_{j+1}$,
provided $c < q$.

Taking into account \eqref{eq:def.psi.j+1}, we have a similar bound for $\supp \psijo_x$.

By \eqref{eq:def.Lam.j+1.F.j+1},  $\lamjo_x - \lamj_{x} = \Wjo_{xx}$, with
$$
\Wjo = \brak{ \Qj\,, \Mjo } + \Dj \Fj  +  (\Mjo)^2 \Lamj  - \Mjo \Lamj \Mjo
$$
(cf. \eqref{eq:W.j+1}), and $\Dj = \Cj - \one$, $\Cj_{yx} = (\phij_x\,, \phij_y)$.
Here, again, we have $\diam \Stoch{\lamjo_x - \lamj_x} = \Ord{cL_j} \le (q - 1)L$ for $c>0$ small enough.
It is readily seen by induction that $x\in\Stoch{\lamj_x} \cap \Stoch{\lamj_x}$, so the required bound on $\Stoch{\lamjo_x}$ follows.

Similar arguments, based on the construction of $\phijo_x$ and $\psijo_x$, prove the bounds $\diam \Stoch{\phijo_x} \le L_{j+1}$
and $\diam \Stoch{\psijo_x} \le L_{j+1}$, if $c>0$ is small enough. Here, too, $x\in\Stoch{\phij_x}\cap\Stoch{\phijo_x}$, $x\in\Stoch{\psij_x}\cap\Stoch{\psijo_x}$.


\par\vskip2mm\noindent
\textbf{Summary of the inductive step.}
For the reader's convenience, we provide below the references to the stages in the proof
where each of the inductive hypotheses (K\ref{lab:K1})--(K\ref{lab:K9}) is proved.
\renewcommand{\arraystretch}{1.0}
$$
\hbox{\begin{tabular}{|l|l|l|l|}
  \hline
 {\small (K\ref{lab:K1})  \, Steps 3, 4, 7}
    & {\small (K\ref{lab:K2})  \, Step 8}
    & {\small (K\ref{lab:K3})  \, Step 4}
\\
  \hline
    {\small (K\ref{lab:K4})  \, Steps 5, 7}
    &  {\small (K\ref{lab:K5}) \,  independent of $j$}
    &  {\small (K\ref{lab:K6})  \,  Step 3, 10}
\\
  \hline
%
   {\small (K\ref{lab:K7}) \, Step 9 }
  &  {\small  (K\ref{lab:K8})  \,  Steps 3, 7, 10}
  &  {\small (K\ref{lab:K9})  \, Steps 4, 6, 7 }

\\
  \hline
\end{tabular}}
$$

\subsection{Proof of Theorem \ref{thm:main.loc}}
\label{sec:proof}

The uniform exponential localization and the unimodality of all eigenfunctions can be established in the same way as in \lcite{C18b}{Section IV}.

\section{Local fluctuations of eigenvalues}
\label{sec:derivatives}

\subsection{General setting and the key lemmata}

The results of this section prepare the ground for an application of Proposition \ref{prop:f.inv} to the proof of Theorem \ref{thm:Main}.
We have to analyze the regularity of $\Th$-probability distributions of finite families of eigenvalues $(\lam_{x_1}\,, \, \ldots \,, \lam_{x_M})$,
starting with $M=2$. This is achieved with the help of the approximate eigenvalues
$\lami_{x_1}$\,, \ldots $\lami_{x_M})$
by induction in $i$. This induction requires also a regularity analysis of $\phii_x$ and $\psii_x\,$.
Lemma \ref{lem:deriv.lamj.L0.L1} sets the base of induction in $i$, and the induction step is covered by Lemma \ref{lem:induction.deriv.lamj}.

The objects $\#^i_x\,$, with "$\#$" standing for "$\lam$", "$\varphi$", or "$\psi$", are considered as functions of the $\Th$-random variables $\th_{n,k}$ (cf. \eqref{eq:def.v}).
For any $x\in\DZ^d$, the objects $\#^i_y$ with $y$ close to $x$ are impacted by independent $\th_{n,k}$,
with $n$ large enough, so the dependence of $\#^i_y$ upon a single variable $\th_{n,k}$
can be studied with the help of the one-parameter families $V(x;t_z) := V(x) + t_z \one_z(x)$, $z\in\DZ^d$.

We use a shortcut $\pttz$ for $\rd/\rd t_z$, but in the particular case $z=0$, we write $\ptt$ ($\equiv \rd/\rd t_0$).
The explicit formulae $\lamz_x = V(x,\om,\th)$, $\Uz = \one$, $\Fz = (\Uz)^{-1} \Psiz = \Psiz$,
$\psiz_x = \sum_{|y-x|=1} \veps\one_y$ show that
\begin{align}
\label{eq:pt.lam0.x.ne.0}
\all z,x\in\DZ^d  \quad \pttz \lamz_x &= \delta_{zx} \,,
\\
\notag
\all z\in\DZ^d \quad \pttz \Uz &= \pttz \Fz = \ptt \Psiz = 0.
\end{align}
It is convenient to introduce the matrices $\Si$, $i\ge 0$, with entries
\bl
\notag
\Si_{yx} :=
\left\{
  \begin{array}{ll}
  (\lami_x - \lami_y)^{-1}  , & \hbox{if $\Mi_{yx} \ne 0$ and $\lami_x \ne \lami_y$;} \\
    0, & \hbox{otherwise.}
  \end{array}
\right.
\el

\ble
\label{lem:deriv.lamj.L0.L1}
Cconsider the functions $t \mapsto \lamz_z\brak{\Lamz(\om,\th) + t\one_0}$, $z\in\ball_{L_0^2}(0)$ (cf. (K\ref{lab:K5}),
and assume that $\abs{ \lamz_x(\om,\th) - \lamz_y(\om,\th)} \ge \delta_0$ for all $x,y\in\ball_{L_0^2}(0)$ with $\abs{y-x}\ge 1$.
Then
\begin{align}
\label{eq:pt.lam.1}
\abs{ \ptt \lamo_x - \ptt \lamz_x } &\le \eps_0^{ \twomi } \,,
\\
\label{eq:pt.phi.1}
\mxnorm{ \ptt\tffi^1_x - \ptt \phiz_x } &\le \eps_0^{ \onemi } \,,
\\
\label{eq:pt.psi.1}
\mxnorm{ \ptt \psio_x } &\le \eps_0^{ \twomi }  \,.
\end{align}

\ele

\ble
\label{lem:induction.deriv.lamj}
Under the assumptions of Lemma \ref{lem:deriv.lamj.L0.L1}, assume in addition that, for some $j\ge 1$ and all $1 \le i \le j$, one has
\begin{align}
\label{eq:pt.lamj}
\abs{ \ptt \lami_x - \ptt \lam^{i-1}_x } &\le \eps_{i-1}^{ \mmi{2} } \,,
\\[5pt]
\label{eq:pt.phij}
\mxnorm{ \ptt \phii_x - \ptt \phi^{i-1}_x } &\le \eps_{i-1}^{ \mmi{ 1 } } \,,
\\[5pt]
\label{eq:pt.psij}
\mxnorm{ \ptt \psii_x } &\le \eps_{i-1}^{  \mmi{2} } \,.
\end{align}
Assume also \eqref{eq:pt.lam.1}--\eqref{eq:pt.phi.1}.
Then  \eqref{eq:pt.lamj}--\eqref{eq:pt.psij} hold true for $i=j+1$.

\ele

Note that by (K\ref{lab:K8}), $\Stoch{ \lami_\bullet} < L_i \,$, thus
%
\bl
\notag
\Stoch{ \#^i_x} \cap \Stoch{ \#^i_y} = \vide \quad \text{ for } \abs{x - y} > 2 L_i \,.
\el
%
Let
\bl
\label{eq:def.hj.x}
\hj(R) :=
\left\lceil  \para{ \ln(R) - \ln (2 L_0)}/ \ln q \right\rceil , \quad R >0\,,
\el
It is readily seen that $\hj(R) := \min \set{ i\ge 0: \, 2 L_i \ge R }$.
To unify notations for the approximate and exact eigenvalues/eigenvectors, let $\#^{\infty}_x $ stand for $\#_x$.
\ble
\label{cor:ptt.remote}
For all $\hj(|x|) \le j \le \infty$, one has
\bl
\notag
\abs{\ptt \lamj_x } & \le \eps_0^{ \frac{ |x|}{2 L_0} } \,,
\quad
\mxnorm{\ptt \phij_x } & \le \eps_0^{ \frac{ |x|}{4 L_0} } \,.
\el
\ele

\proof

For all $0 \le i \le \hj(|x|) -1$, we have, by (K\ref{lab:K8}) and \eqref{eq:def.hj.x}:
\bl
\notag
\all z\in\ball_{L_i}(0)  \;\quad
\Stoch{ \lami_z } \cup \Stoch{ \phii_z } \cup \Stoch{ \psii_z } \subset \ball_{L_i}(z)  \subset \ball_{2L_i}(0) ,
\el
hence
\bl
\notag
\all i\in\lrb{0, \hj(|x|) -1 } \;\;\; \all z \in \DZ^d \mins \ball_{ L_i}(0) \qquad
\ptt \lami_x \,, \; \ptt \phii_x \,, \;  \ptt \psii_x  = 0.
\el
Recalling
%
$\eps_i = \eps_0^{q^i}$ with $q>1$,
it follows that
\bl
\notag
\abs{\ptt \lamj_x } &
\le
\abs{\ptt \lamz_x } + \sum_{1 \le i \le j} \abs{\ptt \lami_x - \ptt \lam^{i-1}_x }
\le \sum_{\hj(|x|) \le i \le j } \eps_i^{ \mmi{2} }
\le \eps_{ \hj(|x|) } \le \eps_0^{ \frac{|x|}{2L_0} } \,.
\el
The bound on $\ptt \phi^{\hj(x)}_x$ is proved in a similar way.
\qed

\subsection{Proof of Lemma \ref{lem:deriv.lamj.L0.L1} }

By \eqref{eq:def.M.j+1} and \eqref{eq:def.Qj}--\eqref{eq:def.tQj}, we have, for $x,y\in\DZ^d$ such that $\Mo_{yx}\ne 0$,
\begin{align}
\notag
\Mo_{yx}
&
= \eps_0 \sum_{z:\, |z-x|=1}  \Sz_{yx} \para{ \one_y \,, \, \one_z }
= \eps_0 \Sz_{yx}\one_{ \set{|y-x|=1 }} \,,
\\
\label{eq:pt.M1}
\ptt \Mo_{yx} &= - \eps_0 \one_{ \set{|y-x|=1 }} (\Sz_{yx})^2 \para{ \ptt \lamz_x - \ptt \lamz_y } \,.
\end{align}
Since $\Uz = \one$, we have $\Dz = 0$.
$\Qz$ is obtained by truncation of
$\tQz = (\Uz)^\top \Psiz = \Const$ (cf. \eqref{eq:def.Qj}), hence $\pt_{t_z} \Qz = 0$ for all $t_z$.
By construction,
\begin{align}
\label{eq:recall.W.F.Lam.1}
\Wo &= [\Qz, \Mo] + \brak{ \Dz\,, \Fz} + (\Mo)^2 \Lamz  - \Mo \Lamz \Mo \,, \quad \Dz = 0,
\\
\notag
\Fo_{yx} &= (1 - \delta_{yx}) \Wo_{yx} \,,
\\
\label{eq:recall.W.F.lam.lam}
\Lamo_{xx} &= \lamo_{x} = \lamz_{x} + \Wo_{xx} \,.
\end{align}

\nblz First, assess $\ptt\lamo_x - \ptt \lamz_x $. By \eqref{eq:recall.W.F.lam.lam} and \eqref{eq:recall.W.F.Lam.1},
\bl
\notag
\ptt \lamo_x - \ptt \lamz_x &= \ptt \Wo_{xx}
= \ptt \para{ [\Qz, \Mo] + (\Mo)^2 \Lamz - \Mo \Lamz \Mo}_{xx}
\el
where, by antisymmetry, $[\Qz, \Mo]_{xx} = 0$, and $\vmnorm{ \Lamz}, \vmnorm{ \ptt\Lamz} \le \Const$.

By assumption, we have $\abs{ \lamz_x - \lamz_y} \ge \delta_0 = \eps_0^{\zplus}$ for all $y$ figuring in non-zero entries of the matrices in \eqref{eq:recall.W.F.Lam.1},
hence
$\delta_0^k = \eps_0^{\zplus}$ and $\eps_0 \delta_0^{-k} \le \eps_0^{\onemi}$, say, for $1\le k \le 2$, which suffices for our purposes. Therefore,
\begin{align}
\notag
\abs{ \ptt \lamo_x - \ptt \lamz_x }
&
\le \abs{ \ptt \para{ (\Mo)^2 \Lamz}_{xx} + \ptt \para{\Mo \Lamz \Mo}_{xx} }
\lea \vmnorm{ \Mo } \vmnorm{ \ptt \Mo } + \vmnorm{ \Mo }^2
\\
\notag
&
\lea \eps_0 \delta_0^{-1} \cdot \eps_0 \delta_0^{-2} + (\eps \delta_0^{-1} )^2 \le \eps_0^{ \twomi }\;\; \text{ (cf. \eqref{eq:norm.Mjo}, \eqref{eq:pt.M1}) }.
\end{align}
\nblz Next, assess $\varphi^1_x - \phiz_x$. By \eqref{eq:def.phijo}, $\tffi^1_x - \phiz_x = \sum_{|y-x|=1} \Mo_{yx}  \phiz_y$ \,,
whence
\bl
\notag
\ptt\tffi^1_x - \ptt \phiz_x &= \ptt\tffi^1_x = \sum\limits_{|y-x|=1} \para{ \ptt \Mo_{yx} } \phiz_y + \sum_{|y-x|=1} \Mo_{yx} \, \ptt \phiz_y
\\
&
= - \eps_0 \sum_{|y-x|=1} (\Sz_{yx})^2 \para{ \ptt \lamz_x - \ptt \lamz_y } \, \phiz_y
\el
with
$\abs{ \ptt \lamz_x - \ptt \lamz_y }  \le \abs{ \ptt \lamz_x }  + \abs{ \ptt \lamz_y } \le 2$ by \eqref{eq:pt.lam0.x.ne.0}. Therefore,
\bl
\label{eq:pt.tffi2.phiz}
\mxnorm{ \ptt\tffi^1_x - \ptt \phiz_x } = \mxnorm{ \ptt\tffi^1_x  } \le \eps_0^{ \mmi{1} } .
\el
To estimate the effect of normalization $\tffio_\bullet \rightsquigarrow \varphi^{1}_\bullet$, one can argue as in the proof of  \eqref{eq:pt.phij}
in the next subsection (cf. \eqref{eq:tffijo-phijo}--\eqref{eq:norm.tffijo.-1}), and conclude that
\bl
\notag
\mxnorm{ \ptt\phi^1_x - \ptt \phiz_x } \le \eps_0^{ \mmi{1} } \,.
\el

\nblz Now, consider the discrepancies. By construction, $\psio_x = \sum_{z \ne x} \Wo_{zx} \tffio_z\,$, so
\bl
\label{eq:pt.psi.1.x}
\ptt \psio_x &= \sum_{z \ne x} \para{ \ptt \Wo_{zx} } \tffio_z  + \sum_{z \ne x} \Wo_{zx}  \ptt \tffio_z \,,
\el
Here, $\vmnorm{\Wjo} \le \eps_0^{\mmi{2}}$ and $\mxnorm{ \ptt \tffio_z} \le \eps_0^{\mmi{1}}$ (cf. \eqref{eq:norm.W.j+1},\eqref{eq:pt.tffi2.phiz}),
so we focus on the first sum in \eqref{eq:pt.psi.1.x}.
By \eqref{eq:recall.W.F.Lam.1},
\bl
\notag
\ptt \Wo &= [ \ptt \Qz, \Mo ] + [ \Qz , \ptt \Mo ] + \ptt \para{ (\Mo)^2 \Lamz + \Mo \Lamz \Mo }
\\
&
= [ \Qz , \ptt \Mo ] + \ptt \para{ (\Mo)^2 \Lamz + \Mo \Lamz \Mo } .
\el
Using the identity \eqref{eq:pt.M1} for $\ptt \Mo_{yx}\,$, we get
\bl
\notag
\vmnorm{ [\Qz, \ptt \Mo] }
&
\le \eps \delta_0^{-1} \vmnorm{ \Qz } \le  \eps^{ \onemi } \cdot C\eps \le \eps^{ \mmi{2} } ,
\\
\vmnorm{  \ptt \para{ (\Mo)^2 \Lamz + \Mo \Lamz \Mo } }
&
\lea \vmnorm{ \Lamz } \vmnorm{ \Mo } \vmnorm{ \ptt \Mo } + 2 \vmnorm{ \Mo }^2 \vmnorm{ \ptt \Lamz }
\\
&
\le \delta_0^{-1} \eps_0 \cdot \eps_0^{\onemi} + C \eps_0^{\mmi{2}} \le \eps_0^{ \mmi{2} } ,
\el
thus
$\vmnorm{ \ptt \Wo } \le \eps_0^{ \mmi{2} } + \eps_0^{ \mmi{2} } \le \eps_0^{ \mmi{2} }$.
Since $\mxnorm{\tffio_z} \le C$, we conclude that
\bl
\notag
\mxnorm{ \ptt \psio_x }
&
\le \eps_0^{ \mmi{2} } + \ord{ \eps_0^{ \mmi{2} }} \le \eps_0^{ \mmi{2} } .
\el

This completes the proof of Lemma \ref{lem:deriv.lamj.L0.L1}.

\subsection{Proof of Lemma \ref{lem:induction.deriv.lamj}}


\nblz We shall need the following auxiliary estimates:
\begin{align}
\label{eq:ptt.M.i+1.M.i}
\vmnorm{ \ptt \Mjo } &\le \eps_j^{ \mmi{1}}  ,
\\
\label{eq:ptt.W.i}
\vmnorm{ \ptt \Wjo } &\le \eps_j^{ \mmi{2}} .
\end{align}
\bre
\label{rem:ind.bounds.lam.phi.psi}
The explicit formulae for $\phiz_\bullet$, $\lamz_\bullet$ and the perturbation estimates \eqref{eq:pt.lamj}--\eqref{eq:pt.phij} with $i\in\lrb{1,j}$,
imply a uniform boundedness of all quantities  $\abs{ \ptt \lami_x }$, $\vmnorm{ \ptt \phii_x }$, $\vmnorm{ \ptt \psii_x }$, $0 \le i \le j$, $x\in\DZ^d$.
\ere

By construction (cf. \eqref{eq:def.lam.j+1}, \eqref{eq:W.j+1}, \eqref{eq:def.M.j+1}, \eqref{eq:def.tQj}--\eqref{eq:def.Qj}):
\begin{align}
\notag
\lamjo_{x} &= \lamj_{x} + \Wjo_{xx} \,,
\\
\notag
\Wjo &= \brak{ \Qj \,, \Mjo } + \Dj \Fj + \para{ \para{\Mjo}^2 \Lamj } - \Mjo \Lamj \Mjo \,,
\\
\label{eq:Mjo}
%
\Mjo_{yx} &= (1 - \delta_{yx}) \; \Si_{yx} \Qj_{yx} \,,
\\
\notag
\Qj_{yx} &= \one_{ \set{ \abs{y-x} \le L_j } } \, \para{ \phij_y\,, \, \psij_x } \,.
\end{align}

\nblz Assess first $\ptt \Mjo_{yx}$.
By \eqref{eq:Mjo}, $\Mjo_{xx} = 0$, so let $1 \le |y - x| \le L_j$, then
\bl
\ptt \Mjo_{yx}
&
= \para{ \ptt \Sj_{yx} } \cdot \Qj_{yx} + \Sj_{yx} \cdot  \ptt \Qj_{yx}
\\
\notag
&
= -\para{ \Sj_{yx} }^2 \, \Qj_{yx}\, (\ptt \lamj_x - \ptt \lamj_y)
+  \Sj_{yx} \para{ \para{ \ptt \, \psij_y \,,\, \phij_x } + \para{ \psij_y \,,\, \ptt \, \phij_x } }
\el
where $\mxnorm{ \psij} \le \eps_j$ by (K\ref{lab:K4}), and $\norm{\phij} = 1$  by (K\ref{lab:K1}), hence
\bl
\notag
\abs{ \Qj_{yx} } \le \abs{ (\psij_y \,, \phij_x) }
\le \eu^{ -m|x-y|} \mxnorm{ \psij} \le \eu^{ -m|x-y|} \eps_j \,.
\el
Further, $\abs{ \lamj_x - \lamj_y }^k \ge \delta_j^k \ge \eps_j^{ \zplus }$ for $y\ne x$ and $k=1,2$, thus
\bl
\notag
\abs{\Sj_{yx}}^2  \, \abs{\Qj_{yx}\, (\ptt \lamj_x - \ptt \lamj_y) }
&
\le C \eu^{-m|x-y|} \, \eps_j^{-\zplus} \, \eps_j \le \eu^{-m|x-y|} \eps_j^{ \mmi{1} } .
\el
On account of the boundedness of $\mxnorm{\pt \phij_x}$ (cf. Remark \ref{rem:ind.bounds.lam.phi.psi}) and $\norm{ \psij_y }_{y} \le \eps_j^{1}$ (cf. \eqref{eq:KLj.norm.psj}),
the estimate \eqref{eq:ptt.M.i+1.M.i} now follows from the inequalities
\begin{align}
\notag
\abs{ \Sj_{yx}  \, \para{ \ptt \, \phij_y \,,\, \psij_x } }
&
\le  \eu^{ -m|x-y| } \, \eps_j^{-\zplus} \, \mxnorm{ \ptt \phij_j } \, \mxnorm{ \psij}
%
%
\le \eu^{ -m|x-y| } \eps_j^{ \mmi{1}}  \,,
\\[3pt]
\notag
\abs{\Sj_{yx}  \, \para{ \phij_y \,,\, \ptt \, \psij_x } } &\le \eu^{ -m|x-y| } \eps_{j-1}^{ \mmi{1}} \,.
\end{align}
%

\nblz Next, let us prove \eqref{eq:ptt.W.i}. We have
\bl
\notag
\vmnorm{ \ptt \Wjo } & \le  \vmnorm{ [\ptt \brak{\Qj , \Mjo} } + \vmnorm{ \ptt \brak{\Dj\,, \Fj}  }
\\
&
+ \vmnorm{ \ptt \bpara{ (\Mjo)^2 \Lamj } } + \vmnorm{ \ptt \bpara{ \Mjo \Lamj \Mjo } } \,
\el
Here, $\Qj  = (\Uj)^{-1} \Psij$, thus
\begin{align}
\notag
\norm{ \ptt \Qj  }
&
\le \norm{ \ptt (\Uj)^{-1} } \, \norm{ \Psij}  + \norm{ (\Uj)^{-1} } \norm{ \ptt \Psij } ,
\\
\notag
\norm{ \ptt (\Uj)^{-1} }  &\le \norm{ (\Uj)^{-1} }^2 \norm{ \ptt \Uj }  \le 4 \norm{ \ptt \Uj }
\end{align}
yielding
\bl
\notag
\norm{ \ptt \Qj  }
&
\le 4 \norm{ \ptt \Uj } \,  \norm{ \Psij}  + 2 \norm{ \ptt \Psij }
\le C  \eps_j \quad \text{(cf. \eqref{eq:KLj.norm.psj})} .
\el
Observe that the uniform in $j$ boundedness of $\norm{ \ptt \Uj }$, where $\Uj_{yx} = \phij_y(x)$, follows from the inductive bound \eqref{eq:pt.phij}, since we have
already derived its variant for $i=j+1$ from the one on the step $j=i$.
Therefore,
\bl
\label{eq:norm.pt.Fj.Mjo}
\vmnorm{\, [ \ptt \Qj , \, \Mjo ] \, }
&
\le C \eps_j^{\mmi{1}} \, \eps_j^{ \mmi{1} }
\le \eps_j^{ \mmi{2} } \,.
\el
Further,
\bl
\notag
\vmnorm{\, [ \Qj , \, \ptt\Mjo ] \, }
&
\le 2 \vmnorm{ \Qj  } \,  \vmnorm{ \ptt \Mjo }
\le 4  \eps_j  \, \vmnorm{ \ptt \Mjo }
\le  \eps_j^{ \mmi{2}}
\el
(cf. \eqref{eq:ptt.M.i+1.M.i}). Next,
\bl
\notag
\vmnorm{ \pt \bpara{  (\Mjo)^2 \Lamj  + \Mjo \Lamj \Mjo } }
&
\lea  \vmnorm{ \Lamj } \vmnorm{\Mjo}  \vmnorm{ \ptt \Mjo }
+ \vmnorm{\ptt \Lamj } \vmnorm{ \Mjo }^2
\\
&
\le \eps_j^{ \mmi{1} } \cdot \eps_j^{ \mmi{1} } +  \eps_j^{ \mmi{2} }
\le \eps_j^{ \mmi{2} } .
\el
By definition, $\Dj = \Cj - \one$ and $\Fj = ( \Uj )^{-1} \Psij$,
so making use of the identity \eqref{eq:C.j.xy.phi.psi}, we obtain in a similar way the bounds
\begin{align}
\label{eq:norm.pt.Dj.Fj}
\vmnorm{ \ptt \Dj } \le \eps_j^{\mmi{1}} ,
\quad
\vmnorm{ \ptt \para{ \Dj \Fj  }} \le \eps_j^{\mmi{2}} .
\end{align}
It follows from \eqref{eq:norm.pt.Fj.Mjo}--\eqref{eq:norm.pt.Dj.Fj} that
\bl
\notag
\abs{ \ptt \Wjo_{yx}  }
\le
\vmnorm{ \ptt \Wjo } \le \eps_j^{ \mmi{2}} .
\el

\nblz For the proof of \eqref{eq:pt.lamj}, notice that, by antisymmetry, $\brak{ \Fj\,, \Mjo }_{xx} = 0$, so
\bl
\notag
\ptt \lamjo_{x} - \ptt\lamj_{x} &= \ptt  \bpara{ \para{\Mjo}^2 \Lamj }_{xx} + \ptt \bpara{ \Mjo \Lamj \Mjo }_{xx}
\el
whence
\bl
\notag
\abs{ \ptt \lamjo_{x} - \ptt \lamj_{x} }
&
\lea \vmnorm{ \Lamj } \;\vmnorm{\Mjo} \; \vmnorm{ \ptt \Mjo }  + 2 \vmnorm{ \Mjo }^2  \vmnorm{\ptt \Lamj }
\\
&
\le \eps_j^{ \mmi{2} } + \eps_j^{ \mmi{2} } \le \eps_j^{ \mmi{2} }
\quad \text{ (cf.  \eqref{eq:ptt.M.i+1.M.i})} .
\el

\nblz \,\textbf{Proof} of the bound \eqref{eq:pt.phij} on $\ptt \phijo_x - \ptt\phij$.
By construction, we have
$\tffijo_x = \phij_x + \sum_{y \ne x} \Mjo_{yx} \, \phij_y$,
so
\bl
\label{eq:pt.tffijo-pttphij}
\vmnorm{ \ptt \tffijo_x - \ptt \phij_x }
&
\le  \sum_{y \ne x} \abs{ \ptt \Mjo_{yx} } \, \mxnorm{ \phij_y } + \sum_{y \ne x} \abs{ \Mjo_{yx}} \, \mxnorm{ \ptt \phij_y }
\\
&
\le \sum_{y \ne x} \para{ \eu^{ -m|x-y| }  \eps_j^{ \mmi{1} } + C \eu^{ -m|x-y| } \eps_j^{ \mmi{1} } } \le \eps_j^{ \mmi{1} }
\quad \text{ (cf.  \eqref{eq:ptt.M.i+1.M.i})} .
\el
%
%
%
Furthermore, it follows from the definition of $\phijo_x$,
\bl
\label{eq:tffijo-phijo}
\phijo_x =  \norm{\tffijo_x}^{-1} \tffijo_x \,,
\quad
\norm{\tffijo_x} = \Bpara{ \sum_y (\tffijo_x(y))^2 }^{1/2} ,
\el
that
\bl
\notag
\ptt \phijo_x
&
= \ptt (\norm{\tffijo_x}^2)^{-\half} \tffijo_x
= - \half \frac{ \ptt (\norm{\tffijo_x}^2)}{\norm{\tffijo_x}^3} \tffijo_x + \frac{1}{\norm{\tffijo_x} } \ptt \tffijo_x
\el
with $\norm{ \tffijo_x} = 1 + \Ord{ \eps_j^{ \mmi{1} }} \equiv \para{1 + \Ord{ \eps_j^{ \mmi{1} }}} \norm{ \phijo_x} $ (cf. \eqref{eq:norm.phijo-phij}), and
\bl
\notag
&\half \, \ptt (\norm{\tffijo_x}^2)
= \sum_z \tffijo_x(z)  \, \ptt \tffijo_x(z)
\\
&
= \sum_z \para{ 1 + \bigOrd{ \eps_{j}^{ \onemi } \, } } \phijo_x(z)  \, \bpara{ 1 + \bigOrd{ \eps_{j}^{ \mmi{ 1 } } \, } } \ptt \phijo_x(z)
\\
&
= \para{ 1 + \bigOrd{ \eps_{j}^{ \onemi } \, } } \ptt \sum_z  (\phijo_x(z) )^2
= \para{ 1 + \bigOrd{ \eps_{j}^{ \onemi } \, } } \ptt \bpara{\norm{ \phijo_x }^2 } .
\el
As mentioned above, we have
$
\norm{ \tffijo_x } = \bpara{ 1 + \bigOrd{\eps_{j+1}^{\mmi{1} } \,} } \norm{ \phijo_x }
$,
whence
\bl
\label{eq:norm.tffijo.-1}
\norm{ \tffijo_x }^{-1} = \bpara{ 1 +  \bigOrd{\eps_{j}^{\mmi{1} } \,} } \norm{ \phijo_x }^{-1} .
\el
Collecting \eqref{eq:pt.tffijo-pttphij}--\eqref{eq:norm.tffijo.-1} along with the estimates of the $\mxnorm{ \tffijo_x - \phijo_x}$
and $\mxnorm{ \ptt \tffijo_x - \ptt \phijo_x}$, we come to the asserted bound \eqref{eq:pt.phij} with $i=j+1$.


\nblz  \,\textbf{Proof} of the bound \eqref{eq:pt.psij} on $\ptt \psijo_x$.
We have by \eqref{eq:H.U.j+1.Psi.j+1} and \eqref{eq:def.F.j+1}
\bl
\notag
\psijo_x &
= \sum_{z \ne x} (1 - \delta_{yx}) \, \Wjo_{yx} \tffijo_z  + \sum_{z \ne x}  \Zjo_{yx} \tffijo_z  \,,
\\[5pt]
\Wjo &= \brak{ \Qj, \Mjo } + \Dj (\Uj)^{-1} \Psij + (\Mjo)^2 \Lamj  + \Mjo \Lamj \Mjo \,,
\el
where $(\Uj)^{-1} \Psij = \Fj$ and $\Zjo$ is defined in \eqref{eq:Z.j+1}.
Consider first $ \ptt \Wjo_{yx} $.
We already have the norm-bounds on $\Qj$, $\Mjo$, $\Dj$, $(\Uj)^{-1}$, $ \Psij$, $\Lamj$, as well as on there derivatives,
and all the terms contributing to $ \ptt \Wjo_{yx} $ are of order of $\eps_j^{\mmi{2}}$. As before, all the bounded factors can be absorbed in $\eps_j^{\mmi{2}}$,
so
%
$
\abs{ \ptt \Wjo_{yx}  } \le \eps_j^{ \mmi{2}} \,.
$
%
A bound on  $\ptt \Zjo$ can be obtained similarly, albeit the calculations are longer,
and the reader can see that it is of order of $\eps_j^{\mmi{3}}$.
Finally,
\bl
\notag
\mxnorm{ \pt \psijo_x }
&
\le \vmnorm{ \ptt \Wjo  } \, \mxnorm{ \tffijo_\bullet } + \vmnorm{ \Wjo  } \, \mxnorm{ \ptt \tffijo_\bullet }
\\
& \;
+ \vmnorm{ \ptt \Zjo  } \, \mxnorm{ \tffijo_\bullet } + \vmnorm{ \Zjo  } \, \mxnorm{ \ptt \tffijo_\bullet }
\le \eps_j^{ \mmi{2}} .
\el
This completes the inductive step and concludes the proof of Lemma \ref{lem:induction.deriv.lamj}.
\qed

\section{Proof of Theorem \ref{thm:Main}}
\label{sec:proof.Main}

We need the following variant of the inverse function theorem (cf. \cite{C21a}).

\bpr
\label{prop:f.inv}
Consider finite-dimensional real normed spaces $(\DX, \normX{\cdot}) $ and
$(\DY, \normY{\cdot})$,
and a mapping $\ff :\DX \to \DY$ differentiable in a ball $\xball_\ell(0)\subset\DX$, $\ell>0$.
Assume that there is an invertible linear mapping $\cA:\,\DX\to\DY$ such that
\beal
\notag
\sup_{\rx \in \xball_\ell(0)} \norm{ \ff '(\rx) - \cA }
\le \eta \le \frac{\mkap}{ \norm{ \cA^{-1} } }  \,, \quad \mkap\in(0, \shalf)\,.
\eeal
Denote
$\ballcA_R(0) := \set{\ry\in\DY:\, \normX{\cA^{-1}\ry} \le R}$, $R\ge 0$.
Then $\ff $ admits a differentiable inverse $\ff ^{-1}:\, \ballcA_{\mkap \ell}(0) \to \xball_\ell(0)$,
and for all $\ry\in\ballcA_{\mkap \ell}(0)$ one has
\be
\label{eq:prop.f.inv.B.f.explicit.again}
\ff ^{-1}(\ry) = \cA^{-1} \ry  + \delta(\ry), \qquad
\normX{\delta(\ry) }
\le 2 \,  \eta\, \norm{ \cA^{-1}} \;  \normX{ \cA^{-1} \ry} \,.
\ee
Furthermore, for any rectangle of the form
%
$
\BLam(\Balpha, \Beps) = I_1 \times \cdots \times I_K
\subset \ball_{\ell/4}(0)\,,
$
%
with
$I_k = [\alpha_k - \eps_k\,, \, \alpha_k + \eps_k]$, $0 < \eps_k \le \quart \ell$,
one has
\bl
\notag
\ff^{-1}\para{ \BLam(\Balpha, \Beps) }
&
\subset \mytimesL_{1\le k \le K} \Bbrak{ \alpha'_k - (1 + \Ord{\eps_k}) \eps_k \,,\, \alpha'_k + (1 + \Ord{\eps_k})\eps_k  }
\el
\epr
%



\proof[Proof of Theorem \ref{thm:Main}]
Fix a set $\cX = (x_1, \ldots,  x_N)\in\cQ^N$ with $\card \cX = N$, and assume that $\cX \subset \ball_R(x_1)$ where $R = \diam \cX$.
Let
\bl
\notag
\hn = \hn(R) &:= \min\brak{ n \ge 1:\, 2^{-n} \le \shalf C_A^{-1} R^{-A} }
\el
(cf. \eqref{eq:Cn.sep.traj} where a similar condition appears),
so that the partition $\cC_\hn$ (see the paragraph following \eqref{eq:def.C.n.k}) separates the phase points
$\set{ T^y\om\,, y\in\ball_R(x_1)}$, including all $T^{x_k}\om$, $k\in\lrb{1,N}$.
It is readily seen that $\hn(R) \asymp \ln R$.

Conditional on $\fB_{\hn}$ (cf. \eqref{eq:def.fB.n}), $\th \mapsto \lam_{x_k}$ are Borel functions of $\th_{\hn,l_k}$, and for each $k\in\lrb{1, K_{\hn}}$,
there is a unique partition element $C_{\hn, l_k}$ with $\supp \mchi_{\hn,l_k} = C_{\hn, l_k} \ni T^{x_k}\om$. For this reason, we "freeze" all $\th_{\hn,l}$ with
$l\not\in \set{l_1, \ldots, l_N}$,
i.e. condition on a sigma-algebra $\fB'_{\hn} \supset \fB_{\hn}$,
and consider $(\th_{\hn,l_1}\,, \, \ldots \th_{\hn,l_N})$ as coordinates $(t_1, \ldots,  t_N)=\Bt$ in $\DR^N$ endowed with the measure
with the density $a_\hn^{-N}\one_{[0, a_{\hn}]^N}(\Bt)$. It follows directly from the results of Section \ref{sec:derivatives} that
the mapping $\ff:\, \Bt \mapsto (\lam_{x_1}(\Bt)\,, \ldots,\lam_{x_N}(\Bt))$, is differentiable, with
\bl
\notag
\ff'(\Bt) = \one + \fD(\Bt),\;\; \norm{\fD(\Bt)} \le \eta := \eps_0^{  \hj(R)} \,,  \;\; \hj(R) \ge 1 \,.
\el
With $\eps_0(\veps)$
small enough, $\ff'(\Bt)$ is invertible, and $\norm{(\ff'(\Bt))^{-1}}\le 1 + \Ord{ \eps_0} \le 3/2$.
Further, let $\cA = \ff'(0)$, then
\bl
\notag
\norm{ \ff'(\Bt) - \one } \le \eta \le \kap/\norm{\cA^{-1}} , \quad \kap = \squart \,.
\el
With $\eta < 1/16$, we have (cf. \eqref{eq:prop.f.inv.B.f.explicit.again})
$\norm{\ff^{-1}(\Bt) - \cA^{-1} \Bt} \le \half \norm{\Bt}$,
thus the inverse image $\ff^{-1}(I'_1\times \cdots \times I'_N)$ is covered by a rectangle $I''_1 \times \cdots \times I''_N$ with $|I''_k| \le 2|I'_k|\le 4|I_k|$,
$k\in \lrb{1,N}$.
Concluding,
\bl
\notag
\prth{ \all k\in\lrb{1, N} \;\; \lam_{x_{k}} \in I'_k  }  \lea a^{-N}_\hn \abs{I_1} \cdots \abs{I_N} \,,
\el
with $a_{\hn(R)} \ge \eu^{-C \ln^2(R)}$, so the claim follows.
\,
\qed

\begin{appendices}

\section{Localization in lattice subsets}
\label{app:loc.simple.sets}

A thorough inspection of the scale induction evidences that it can be carried out, \emph{mutatis mutandis}, in certain finite or infinite
lattice subsets $\cQ \subset\DZ^d$
with a sufficiently simple geometry of the boundary $\pt \cQ$. By way of example, we describe here one class of such subsets.

Call a connected subset $\cQ\subsetneq \DZ^d$ \emph{simple} if $\pt\cQ$ is contained in a finite union of convex subsets of lower-dimensional coordinate lattices
and of their shifts.
For example, one can consider intersections of a finite number of half-lattices,
\be
\notag
\cQ = \set{x\in\DZ^d:\; s_i (x_i - c_i) \ge 0,\, 1 \le i \le k}\,, \;\; c_i\in\DZ, \;\;  s_i = \pm 1,
\ee
including the lattice rectangles $\lrb{a_1\,, b_1} \times \cdots \times \lrb{a_d\,, b_d}$.

\btm
\label{thm:loc.simple.sets}
The assertions of Theorem \ref{thm:main.loc} remain valid for a self-adjoint restriction of the Hamiltonian \eqref{eq:def.H}
to a simple subset $\cQ\subsetneq \DZ^d$.
\etm

To illustrate the main idea of such an extension, consider the simplest nontrivial case where $\cQ = \DZ_+$.
On the step $i=0$, both $\lamz_x$ and $\phiz_x$ have the same form in the entire $\cQ$, up to a shift.
The discrepancies are sensitive to the presence of the boundary point $0$, but they affect the approximate eigenvalues and eigenvectors only on the subsequent steps.
Further, for a fixed, $i\ge 1$, only a finite number of positions $x\in \DZ_+$ carry the objects $\#^i_x$ that cannot be derived by covariance from $\#^i_0$:
those with $\abs{x - 0}\le \Ord{L_i}$. In higher dimension $d>1$, their number is bounded by $\Ord{L_i^d}$.

These combinatorial bounds are to be used only on Step 9 where we eliminate the unwanted values $\th\in\Th$ providing abnormally small spacings
$\abs{\lamjo_x - \lamjo_y}$ with $\abs{x - y}\le \Ord{L_{j+1}}$. A direct inspection of the $\Th$-probability estimates used on Step 9 evidences
that a finite number of exceptional pairs $(x,y)$, polynomially bounded in $L_i$, does not destroy the principal mesure-theoretic bounds, and so
the entire induction in $i\in\DN$ can go through.

\section{Estimates of higher derivatives}

As the reader has undoubtedly noticed in Section \ref{sec:derivatives}, nothing prevents one from deriving $\ptt \lami_\bullet$,
$\ptt \lami_\bullet$ and $\ptt \lami_\bullet$ one more time, and by recurrence, a finite number of times. However, it is to be reminded that the
simplified notation $\eps_j^{\mpmi{b}}$ can absorb various bounded factors and the inverses $(\eps_j^{\mpi{0}})^{-1}$ only a finite number of times.
Eventually, the explicitly written powers $(\eps_j^{-c})^k$ can overcome and destroy the small factors $\eps_j^{b}$, $b>0$, unless $c>0$
is made smaller, depending upon $k$. In turn, to keep the powers $\eps_j^c$ with a smaller $c>0$ meaningful, one has to decrease $\eps_0(\veps) = \veps^{1/4}$,
hence decrease $\veps>0$. With these observations in mind, we come to the following generalization of Lemma \ref{lem:deriv.lamj.L0.L1}
and Lemma \ref{lem:induction.deriv.lamj}.

\bpr
For any $N\ge 2$, there exists $\veps(N)>0$ such that, for any $\veps \in\bpara{ 0, \veps(N) }$,
the derivatives of orders $r=1, \ldots, N$ of $\lami_x$, $\phii_x$ and $\psii_x$, $0 \le i \le \infty$, are well-defined,
and the bounds of the form \eqref{eq:pt.lam.1}--\eqref{eq:pt.psi.1} and \eqref{eq:pt.lamj}--\eqref{eq:pt.psij}
remain valid with $\pt_t$ replaced with $\pt_t^r$, $r=1, \ldots, N$.
\epr

\end{appendices}



\end{document}